\documentclass[preprint,12pt]{elsarticle}

\usepackage{amssymb, bm}
\usepackage{amsthm, amsmath}

\usepackage{graphicx}
\usepackage{caption}
\usepackage{subcaption}
\usepackage{multirow}
\usepackage{mathtools,xparse}

\makeatletter
\newenvironment{smallermatrix}[1][c]
{\null\,\vcenter\bgroup
  \Let@\restore@math@cr\default@tag
  \baselineskip0pt \lineskip0.4pt \lineskiplimit0pt
  \ialign\bgroup\if#1l\else\hfil\fi$\m@th\scriptstyle##$\if#1r\else\hfil\fi&&\thickspace\hfil
  $\m@th\scriptstyle##$\hfil\crcr
}{%
  \crcr\egroup\egroup\,%
}
\makeatother

\NewDocumentCommand{\ts}{O{c} e{^?_}}{
  \begin{smallermatrix}[#1]
  \mathstrut\IfValueT{#2}{#2} \\
  \mathstrut\IfValueT{#3}{#3} \\
  \mathstrut\IfValueT{#4}{#4}
  \end{smallermatrix}%
}

\journal{Arxiv}

\begin{document}

\begin{frontmatter}



\title{Proper Orthogonal Descriptors for Efficient and Accurate Interatomic Potentials}




\author[inst2]{Ngoc-Cuong Nguyen}
\author[inst3]{Andrew Rohskopf}

\affiliation[inst2]{organization={Center for Computational Engineering, Department of Aeronautics and Astronautics, Massachusetts Institute of Technology},
            addressline={77 Massachusetts
Avenue}, 
            city={Cambridge},
            postcode={02139}, 
            state={MA},
            country={USA}}
            
\affiliation[inst3]{organization={Department of Mechanical Engineering, Massachusetts Institute of Technology},
            addressline={77 Massachusetts
Avenue}, 
            city={Cambridge},
            postcode={02139}, 
            state={MA},
            country={USA}}

\begin{abstract}
We present the proper orthogonal descriptors for efficient and accuracy  representation of the potential energy surface. The potential energy surface is represented as a many-body expansion of parametrized potentials in which the potentials are functions of atom positions and parameters. The Karhunen-Lo\`eve (KL)  expansion is employed to decompose the parametrized potentials into a set of proper orthogonal descriptors (PODs).  Because of the rapid convergence of the KL expansion, relevant snapshots can be sampled exhaustively to  represent the  atomic neighborhood environment accurately with a small number of descriptors. The proper orthogonal descriptors are used to develop interatomic potentials by using a linear expansion of the descriptors and determining the expansion coefficients from a weighted least-squares regression against a density functional theory (DFT) training set. We present a comprehensive evaluation of the POD potentials on previously published DFT data sets comprising Li, Mo, Cu, Ni, Si, Ge, and Ta elements. The data sets represent a diverse pool of metals, transition metals, and semiconductors. The accuracy of the POD potentials are comparable  to that of state-of-the-art machine learning potentials such as the spectral neighbor analysis potential (SNAP) and the atomic cluster expansion (ACE).
\end{abstract}






\end{frontmatter}


\section{Introduction}
\label{sec:intro}

Decomposing complex systems into a hierarchy of interacting subsytems at different scales is a recurring theme in physics. The approach applied to atomic systems lends itself naturally to learning atomic interactions. For example, to calculate the total energy, one might just sum up independent contributions from each of its neighbors. Specifically, the potential energy $E$ is constructed as a sum of local atomic energies $E_i$ of all atoms $i$ in the system
\begin{equation}
\label{eq0}
E(\bm r_1,\ldots, \bm r_N) = \sum_{i = 1}^N E_i(\bm r_1,\ldots, \bm r_N) ,
\end{equation}
where $\bm r_i$ are Cartesian coordinates of atoms $i$ in the system. Although the local energies $E_i$ are generally functions in  $3N$-dimensional space, they are often expressed in much lower dimensional spaces by  limiting the atomic interaction to local atomic environments around the  atoms $i$. Typically, the local energies take the form
\begin{equation}
\label{eq3b}
E_i = \sum_{m=1}^M \Psi_m(\bm q_i, \bm c_m) , \quad i = 1,\ldots, N ,
\end{equation}
where $\bm q_i$ are the local coordinates attached to each atom $i$, $\{\bm c_m\}_{m=1}^M$ are sets of parameters for  potential functions $\{\Psi_m(\bm \cdot, \bm \cdot)\}_{m=1}^M$, which together encode local atomic environments around atoms $i$. The local coordinates $\bm q_i$ include coordinates of the atom $i$ and those of the neighbors within a cutoff distance  from the atom $i$. Since the number of neighbors around any atom $i$ can often be orders of magnitude less than $N$, the expansion (\ref{eq3b}) allows for substantial dimensionality reduction.



Over the years, a large number of empirical potentials have been developed to provide analytical forms of the potential functions $\Psi_m$. Empirical potentials are usually derived from physical insights of electronic structure theories. Despite many years of efforts devoted to the development of empirical potentials, open problems still need to be addressed. Empirical potentials often suffer from not being able to reproduce properties of a system under different temperature, pressure and environmental conditions. More often than not, empirical potentials  are  recalibrated by reparameterizing the existing set of parameters or extending the energy formula by adding terms and then parameterizing the entire potential anew. This results in many different sets of parameters for widely-used empirical potentials such as EAM \cite{Daw1984},  Stillinger-Weber \cite{Stillinger1985}, Tersoff \cite{Tersoff1988}, EDIP \cite{Bazant1997}, REBO \cite{Brenner2002}, ReaxFF \cite{VanDuin2001}. Empirical potentials and their specific parametrization should be applied to problems for which  they are known to produce consistently accurate results. Developing accurate interatomic potentials for a wide variety of systems is still an active and important area of research 

In recent years, a significant trend has emerged in the form of machine learning potentials, where the potential energy surface is described as a function of local environment descriptors that are invariant to translation, rotation, and permutation of atoms in the system. Examples of such potentials include the high-dimensional neural network potential (NNP) \cite{Behler2007,Behler2011,Behler2014}, the
Gaussian approximation potential (GAP) \cite{Bartok2010,Fujikake2018,Szlachta2014}, the spectral neighbor analysis potential (SNAP) \cite{Thompson2015,Wood2018}, moment tensor potentials (MTP)  \cite{Novoselov2019,Shapeev2016}, and the atomic cluster expansion \cite{Drautz2019,Drautz2020,DUSSON2022110946}. Typical approach to training such potentials involves the generation of a sufficiently large and diverse data set of atomic configurations with corresponding energies, forces and stresses from DFT
calculations, which are then used in the training of ML potentials based on one or several target metrics, such as minimizing the mean absolute or squared errors in predicted energies, forces, stresses, or derived properties (e.g., elastic constants).  A recent work \cite{Zuo2020} presents a comprehensive evaluation of ML potentials based on four local environment descriptors - atom-centered symmetry functions (ACSF), smooth overlap of atomic positions (SOAP), the SNAP bispectrum components, and moment tensors. It is shown in \cite{Zuo2020}  that ML potentials provide  substantial improvement over  EAM and Tersoff potentials in predicting energies and forces across diverse chemistries and configurations.


The  descriptors play a central role in the accurate representation of the potential energy. They must meet several requirements.  First and foremost, the descriptors must be invariant with respect to permutation, rotation, translation, and reflection of atoms in the system. Second, since the transformation from atom coordinates onto the descriptors has to be carried out for every atom, the evaluation should be fast, and the descriptors need to be differentiable with respect to the atomic positions to enable the calculation of analytic gradients for the forces. Third, in order to obtain accurate ML potentials, the descriptors must provide a  detailed structural description of the local atomic environment.  The challenge of finding suitable descriptors for ML potentials was recognized in earlier work \cite{Behler2007}.  Initially, the lack of suitable descriptors was the main obstacle to construct ML potentials for high-dimensional systems because multidimensional functions were required to fit the many-body interactions in electronic structure calculations with high precision. In recent years, significant progress has been made in developing  a wide variety of descriptors for ML potentials. We refer to \cite{Bartok2013,Behler2016,DUSSON2022110946} for a comprehensive survey of methods to transform the coordinates of atoms onto a suitable set of descriptors. 

In this paper, we present a new method to represent atomic neighborhood environments with the so-called proper orthogonal descriptors. The method is motivated by the reduced basis method \cite{barrault04:_empir_inter_method,Boyaval2009a,Grepl2007a,ARCME,Nguyen2008} for parametrized partial differential equations. In spirit of the reduced basis method, we view the potential energy surface as a manifold of potential energies over parameters. Instead of calibrating or optimizing the parameters, we aim to construct an efficient and accurate approximation of the manifold of potential energies, thereby providing a rich and diverse representation of atomic interactions. To this end, we employ the Karhunen-Lo\`eve (KL) expansion \cite{sirovich87:_turbul_dynam_coher_struc_part,everson95karhunenloeve,willcox02:_balanced_pod} to generate orthogonal basis functions that optimally approximate the manifold of potential energies. These basis functions are then transformed onto a set of proper orthogonal descriptors (PODs) to represent atomic neighborhood environments.  Owing to the exponential convergence of the KL expansion, accurate  representation of local atomic environments can be achieved with a small number of descriptors. The proper orthogonal descriptors are used to develop ML potentials by using a linear expansion of the descriptors and determining the expansion coefficients from a weighted least-squares regression against DFT training set. We assess the POD potentials on diverse data sets of Li, Mo, Cu, Ni, Si, Ge, and Ta elements, and compare their performance to that of other linear models such as SNAP and ACE potentials.

The paper is organized as follows. In Section 2, we introduce proper orthogonal descriptors. In Section 3, we discuss the connection of our work with previous work. In Section 4, we describe the implementation of the proposed descriptors and the potential fitting. In Section 5, we present numerical results to assess the performance of the proposed method. Finally, in Section 6, we make a number of concluding remarks on the results as well as future work.

\section{Proper Orthogonal Descriptors}

\subsection{Parametrized potential energy surface}

Let $\bm r_n \in \Omega$ be a position vector of an atom $n$ in a physical domain $\Omega \in \mathbb{R}^3$. We consider a system of $N$ atoms with $N$ position vectors $\bm R = (\bm r_1, \bm r_2, \ldots, \bm r_N) \in \mathbb{R}^{3N}$. The potential energy surface (PES) of the system of $N$ atoms can be expressed as a many-atom expansion of the form
\begin{equation}
\begin{split}
E(\bm R, \bm \eta, \bm \mu) \ = \ & V^{(0)} + \sum_{i} V^{(1)}(\bm r_i, \bm \mu^{(1)} ) + \frac12 \sum_{i,j} V^{(2)}(\bm r_i, \bm r_j, \bm \eta^{(2)}, \bm \mu^{(2)})  \\
& + \frac16 \sum_{i,j,k} V^{(3)}(\bm r_i, \bm r_j, \bm r_k, \bm \eta^{(3)}, \bm \mu^{(3)}) + \ldots 
\end{split}
\end{equation}
where $V^{(0)}$ is a constant energy offset, $V^{(1)}$ is the one-body potential often used for representing external field or energy of isolated elements, and the higher-body potentials $V^{(2)}, V^{(3)}, \ldots$ are symmetric, uniquely defined, and zero if two or more indices take identical values. The superscript on each potential denotes its body order. Note that each $q$-body potential $V^{(q)}$  depends on $\bm \mu^{(q)}$ and $\bm \eta^{(q)}$, which are a set of parameters to fit the potential and a set of hyperparameters, respectively. Then $\bm \mu$ is a collection of all potential parameters $\bm \mu^{(1)}$, $\bm \mu^{(2)}$, $\bm \mu^{(3)}, \ldots$, and $\bm \eta$ is a collection of all hyperparameters $\bm \eta^{(2)}$, $\bm \eta^{(3)}$, etc. The hyperparameters include inner and outer cut-off distances. Typically, lower-body potentials have longer cut-off distances than higher-body potentials.  Note that the one-body potential $V^{(1)}$ does not have hyperparameters since there is only one central atom $i$ in the one-body interaction. 


In the absence of external fields, the PES should not depend on the absolute position of atoms, but only on the relative positions. This means that the PES can be rewritten as a function of interatomic distances and angles between the bonds as 
\begin{equation}
\label{eq2}
\begin{split}
E(\bm R, \bm \eta, \bm \mu) \ = \ & V^{(0)} + \sum_{i} V^{(1)}( \bm \mu^{(1)}) + \frac12 \sum_{i,j} V^{(2)}( r_{ij}, \bm \eta^{(2)}, \bm \mu^{(2)})  \\
& + \frac16 \sum_{i,j,k} V^{(3)}(r_{ij}, r_{ik}, \theta_{ijk}, \bm \eta^{(3)}, \bm \mu^{(3)}) + \ldots 
\end{split}
\end{equation}
where each  potential is a function of interatomic distances $r_{ij} = |\bm r_i - \bm r_j|$, bond angles $\theta_{ijk}$, potential parameters, and the hyperparameters.  A wide variety of interatomic potentials such as Lennard-Jones potential, Morse potential, Stillinger-Weber potential, angle potentials, and dihedral potentials have the form (\ref{eq2}).

In other interatomic potentials such as EAM and Tersoff potentials, the many-body interactions are embedded into the terms of a pair potential in which the nature of the interaction is modified by the local environment of the atom via the bond order parameter, coordination number, or electron density. For the EAM potential, the energy is a nonlinear function of a sum of functions of the interatomic distance. The Tersoff potential includes an explicit dependence on bond angles to describe directional bond formation, so that the energy is written as a nonlinear function of a sum of functions of the interatomic distance and bond angle.


Interatomic potentials rely on parameters to learn relationship between atomic environments and interactions.  Since interatomic potentials are approximations by nature, their parameters need to be set to some reference values or fitted against experimental and/or DFT data sets by necessity. In simple potentials such as the Lennard-Jones and Morse potential, the parameters can be set to match the equilibrium bond length and bond strength of a dimer molecule or the surface energy of a solid. Many-body potentials often contain many parameters with limited interpretability and need to be optimized by fitting their parameters against a larger set of data. Typically, potential fitting involves solving a regression problem to find optimal parameters, $\bm \mu^*$, which  minimize a certain loss function of the predicted quantities and data. The accuracy of a fitted potential at different conditions other than those used in the fitting procedure is often measured in terms of transferability of the potential. It is desirable to make the potential transferable, so that it can describe  properties that are different from those it was fitted to. 
 Since the fitted potential depends on the data set used to fit it, different data sets will yield different optimal parameters and thus different fitted potentials. When fitting the same functional form on $Q$ different data sets, we would obtain $Q$ different optimized potentials, $E(\bm R, \bm \eta, \bm \mu_q^*), 1 \le q \le Q$. These optimized potentials are typically intended to predict properties that are similar to those they are fitted to. Inaccurate predictions can occur when they are used to predict out-of-fitting properties. Moreover, the hyperparameters are left to be defined by users of the potentials.

Instead of finding optimal parameters for the PES (\ref{eq2}), we draw inspiration from the reduced basis method \cite{barrault04:_empir_inter_method,Boyaval2009a,Grepl2007a,ARCME,Nguyen2008} for parametrized partial differential equations and propose to view the parametrized PES as a parametric manifold of potential energies 
\begin{equation}
\mathcal{M} = \{E(\bm R, \bm \eta, \bm \mu) \ | \  \bm \mu \in \Omega^{\bm \mu} \}    
\end{equation}
where $\Omega^{\bm \mu}$ is a parameter domain in which $\bm \mu$ resides. The parametric manifold $\mathcal{M}$ contains potential energy surfaces for all values of $\bm \mu \in \Omega^{\bm \mu}$.  
Therefore, the parametric manifold yields a much richer and  more transferable atomic representation than any particular individual PES $E(\bm R, \bm \eta, \bm \mu^*)$.
Our goal in this section is to construct an efficient and accurate approximation to the parametric manifold $\mathcal{M}$. To this end, we apply the KL expansion to snapshots (different values of parameters $\bm \mu$) of the parametric manifold to obtain a set of proper orthogonal descriptors (PODs). The descriptors can then be employed in a fitting scheme, such as linear regression, to develop accurate and transferable interatomic potentials. The first step to crafting the PODs is to define a parametric manifold and its domain of parameters.




\subsection{Two-body proper orthogonal descriptors}

To define a parametric manifold for two-body interactions, we adopt the usual assumption that the direct interaction between two atoms vanishes smoothly when their distance is greater than the outer cutoff distance $r_{\rm max}$. Furthermore, we assume that two atoms can not get closer than the inner cutoff distance $r_{\rm min}$ due to the Pauli repulsion  principle. Letting $r \in (r_{\min}, r_{\max})$, we introduce the following parametrized radial functions
\begin{equation}
\label{eq3}
\phi(r, r_{\rm min}, r_{\rm max}, \alpha, \beta)  = \frac{\sin (\alpha \pi x) }{r - r_{\rm min}}, \qquad  \varphi(r, \gamma)  = \frac{1}{r^\gamma} ,    
\end{equation}
where the scaled distance function $x$ is defined below to enrich the two-body manifold   
\begin{equation}
x(r, r_{\rm min}, r_{\rm max}, \beta) = \frac{e^{-\beta(r - r_{\rm min})/(r_{\rm max} - r_{\rm min})} - 1}{e^{-\beta} - 1} .
\end{equation}
We introduce the following function as a convex combination of the two functions in (\ref{eq3})
\begin{equation}
\psi(r, r_{\rm min}, r_{\rm max}, \alpha, \beta, \gamma, \kappa)  = \kappa \phi(r, r_{\rm min}, r_{\rm max}, \alpha, \beta) + (1- \kappa)  \varphi(r, \gamma) .
\end{equation}
We see that $\psi$ is a function of distance $r$, cut-off distances $r_{\min}$ and $r_{\max}$, and parameters $\alpha, \beta, \gamma, \kappa$. Together these parameters allow the function $\psi$ to characterize a diverse spectrum of two-body interactions within the cut-off interval $(r_{\min}, r_{\max})$. 

\begin{figure}[htbp]
\centering
\includegraphics[width=0.65\textwidth]{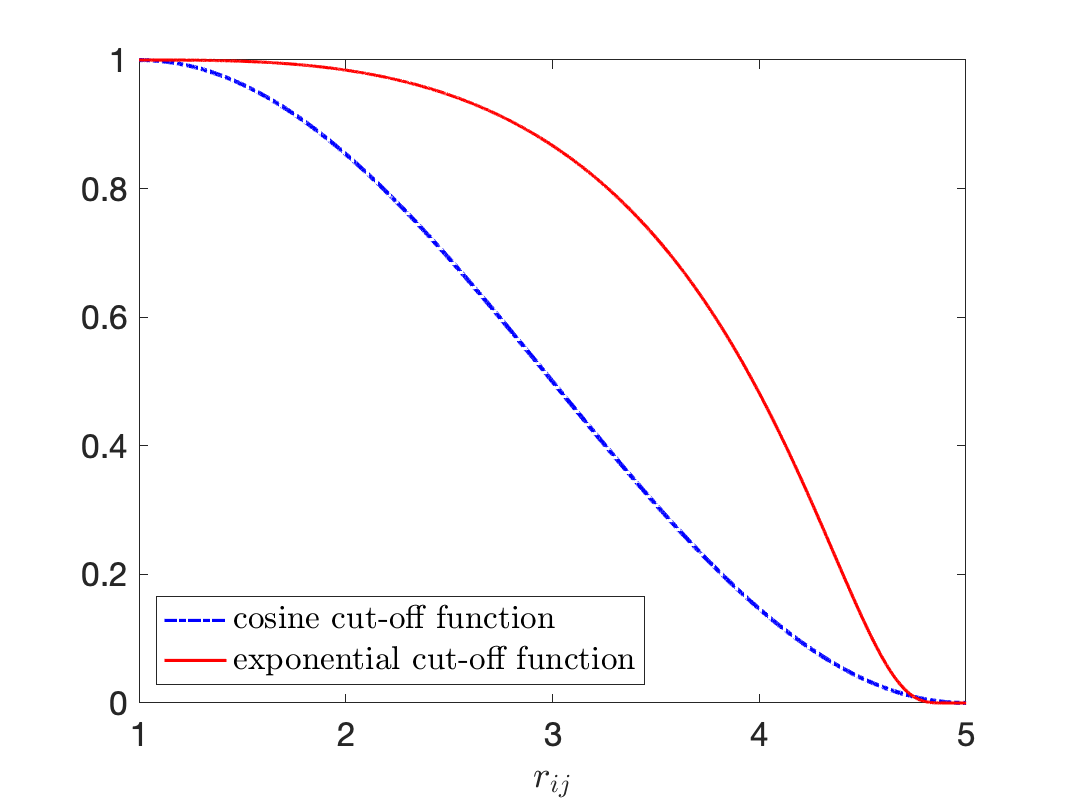}
\caption{Cut-off functions (\ref{eq7})  and (\ref{eq8})  as a function of $r_{ij}$ for $r_{\min} = 1$\AA \ and $r_{\max} = 5$\AA.}
\label{fig:cutoff}	
\end{figure}

Next, the two-body potential is defined as follows
\begin{equation}
\label{eq6}
V^{(2)}(r_{ij}, \bm \eta^{(2)}, \bm \mu^{(2)})  = f_{\rm c}(r_{ij}, \bm \eta^{(2)}) \psi(r_{ij}, \bm \eta^{(2)}, \bm \mu^{(2)})
\end{equation}
where $\eta^{(2)}_1 = r_{\min}, \eta^{(2)}_2 = r_{\max}, \mu_1^{(2)} = \alpha, \mu_2^{(2)} = \beta, \mu_3^{(2)} = \gamma$, and $\mu_4^{(2)} = \kappa$. Here the cut-off function $f_{\rm c}(r_{ij}, \bm \eta^{(2)})$ is used to ensure the smooth vanishing of the potential and its derivative for $r_{ij} \ge r_{\max}$. The following cut-off function has been widely used in many interatomic potentials:
\begin{equation}
\label{eq7}
 f_{\rm c}(r_{ij},  r_{\rm min}, r_{\rm max})  = 0.5 \left( 1 + \cos \frac{\pi(r_{ij} - r_{\min})}{r_{\max} - r_{\min}} \right) .
\end{equation}
Here we propose a new cut-off function 
\begin{equation}
\label{eq8}
 f_{\rm c}(r_{ij},  r_{\rm min}, r_{\rm max})  =  \exp \left(1 -\frac{1}{\sqrt{\left(1 - \frac{(r-r_{\min})^3}{(r_{\max} - r_{\min})^3} \right)^2 + \epsilon}} \right)
\end{equation}
where $\epsilon = 10^{-6}$. As shown in Figure \ref{fig:cutoff}, the new cut-off function (\ref{eq8}) has considerably higher values than (\ref{eq7}) on almost the entire interaction distance. As a result, (\ref{eq8}) can enable the potentials to capture more atomic interactions than (\ref{eq7}). We empirically observe that the cut-off function (\ref{eq8}) often yields more accurate results in terms of both energy and force predictions than (\ref{eq7}).

\begin{figure}[h]
	\centering
	\begin{subfigure}[b]{0.49\textwidth}
		\centering
		\includegraphics[width=\textwidth]{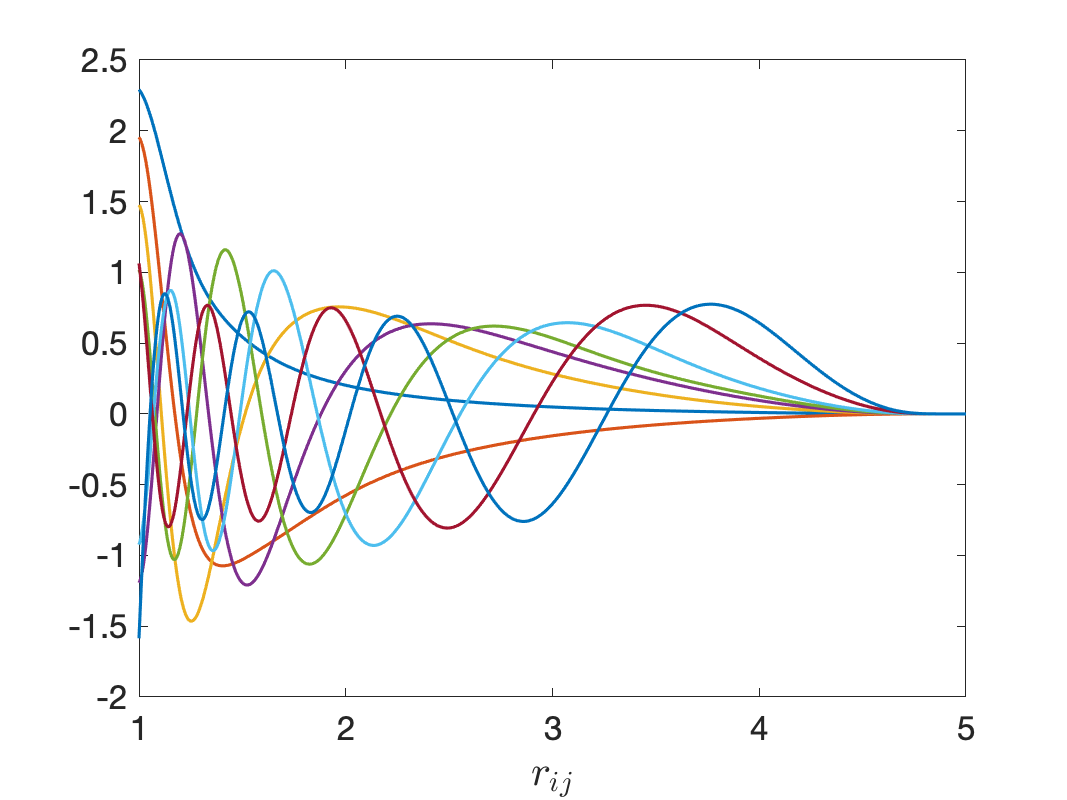}
		\caption{Orthogonal basis functions.}
		\label{fig:mesh}
	\end{subfigure}
	\hfill
	\begin{subfigure}[b]{0.49\textwidth}
		\centering
		\includegraphics[width=\textwidth]{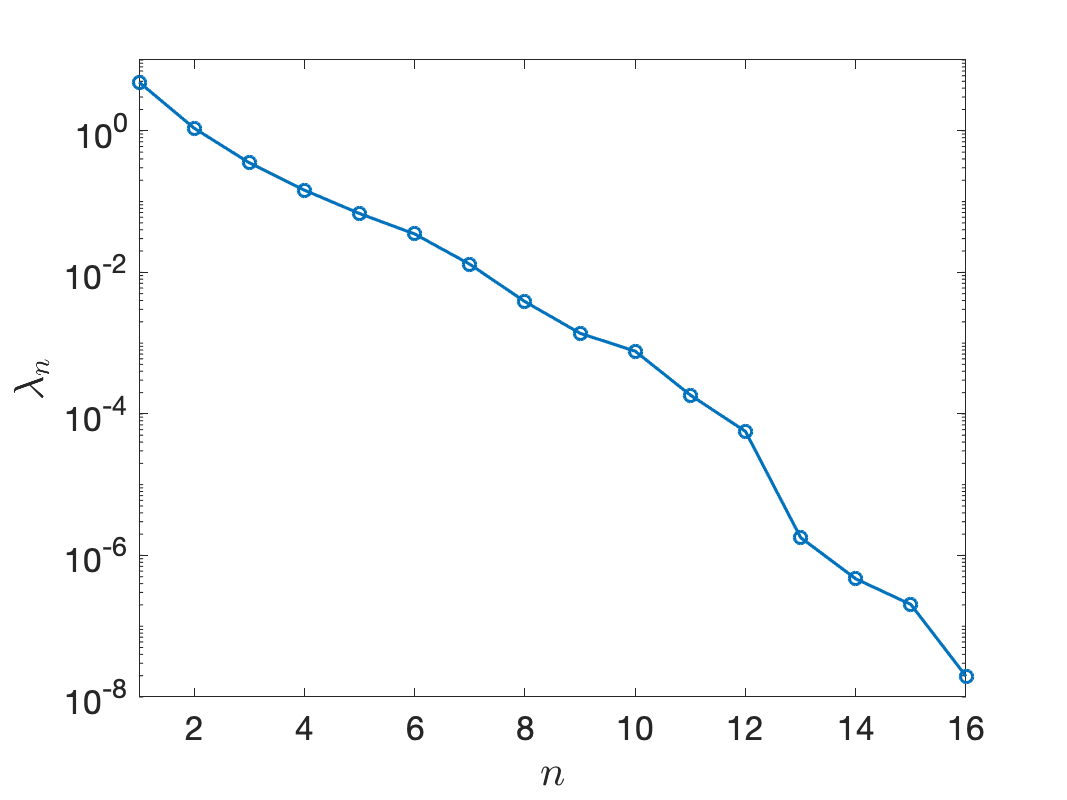}
		\caption{Eigenvalues.}
		\label{fig:displacement}
	\end{subfigure}
	\caption{The first 8 orthogonal basis functions and first 16 eigenvalues for the KL expansion of the two-body parametrized potential (\ref{eq6}).}
	\label{fig:basis}
\end{figure}

Based on the two-body parametrized potential (\ref{eq6}), we construct the proper orthogonal descriptors as follows. We assume that we are given $L$ parameter tuples $\bm \mu^{(2)}_\ell, 1 \le \ell \le L$. The selection of these parameters is explained in Section 4. We introduce the set of  snapshots on $(r_{\min}, r_{\max})$ as
\begin{equation}
\label{eq11}
\xi_\ell(r_{ij}, \bm \eta^{(2)}) =  V^{(2)}(r_{ij}, \bm \eta^{(2)}, \bm \mu^{(2)}_\ell),  \quad \ell = 1, \ldots, L .
\end{equation}
We note here that these snapshots are not evaluated on any particular set of atomic configurations in a training set; the snapshots are calculated on the interval $r_{\min} \leq r_{ij} \leq r_{\max}$ to ensure adequate sampling of the PES for different parameters and hyperparameters. We assume that the number of snapshots $L$ is sufficiently large to approximate the parametric manifold of potential energies well. Typically, we choose $L$ between $10$ and $30$. We employ the Karhunen-Lo\`eve (KL) expansion~\cite{sirovich87:_turbul_dynam_coher_struc_part} to generate an orthogonal basis set which is known to be optimal for representation of the snapshot family $\{\xi_\ell\}_{\ell=1}^L$. The details of the KL expansion are given in~\ref{sec::appendix} for reference. In particular, the two-body  orthogonal basis functions are computed as follows
\begin{equation}
\label{eq12}
U^{(2)}_k(r_{ij}, \bm \eta^{(2)}) = \sum_{\ell = 1}^L a_{\ell k} \,  \xi_\ell(r_{ij}, \bm \eta^{(2)}), \qquad k = 1, \ldots, K , 
\end{equation}
where the matrix components $a_{\ell k}$ consist of eigenvector components of the eigenvalue problem (\ref{eq3a:7}), which finds the components in the space of snapshots with the largest spatial variability between snapshots. The number of basis functions $K$ is less than $L$, and consist of the largest eigenvalues. Figure \ref{fig:basis} shows the orthogonal basis functions as well as the eigenvalues for $r_{\rm min} = 1$\AA \ and $r_{\max} = 5$\AA, for a family of 40 snapshots. We observe the exponential decay of the eigenvalues, which is why we can take $K < L$. For example, 10 basis functions ($K=10$) capture more than $99.99 \%$ the energy content of the family of 40 snapshots ($L=40$).

\begin{figure}[htbp]
\centering
\includegraphics[width=1.0\textwidth]{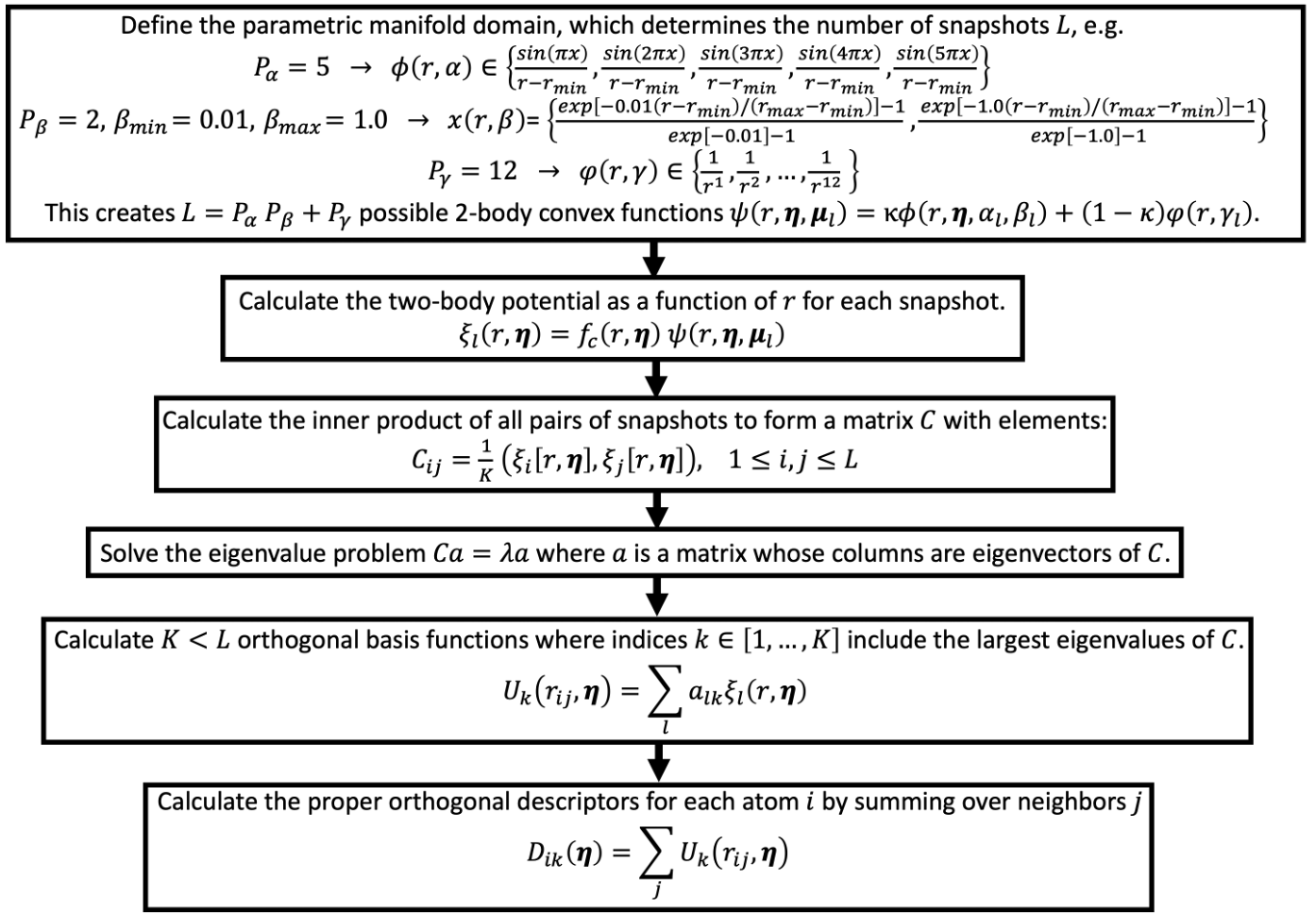}
\caption{Procedure for making two-body proper orthogonal descriptors. First we define the parametric manifold and its parameter domain by specifying the functional forms and their sets of parameters. These sets are created by specifying the bounds $P_\alpha$ and $P_\gamma$ for which parameters $\alpha \in [1, P_\alpha]$ and $\gamma \in [1, P_\gamma]$ take integer values. The parameter $\beta$ can take $P_\beta$ values in the range $[\beta_{min}, \beta_{max}]$. We let $\kappa$ be 0 or 1, thus resulting in $L=P_\alpha P_\beta + P_\gamma$ sets of two-body convex functions $\psi = \kappa \phi + (1-\kappa) \varphi$, from which the snapshots $\xi_l$ are formed. Next, we calculate the spatial inner product between all pairs of snapshots to form a matrix $C$, which may be thought of as a covariance matrix that measures spatial variability between snapshots. The largest eigenvalues of this matrix are associated with eigenvectors, or components in the space of snapshots, pointing in the direction of greatest variability between snapshots. We therefore define the two-body orthogonal basis functions $U_k$ as expansions in terms of the snapshots $\xi_l$ modulated by their eigenvector components $a_{lk}$.}
\label{fig:pod_procedure}	
\end{figure}

Finally, the two-body proper orthogonal descriptors at each atom $i$ are computed by summing the orthogonal basis functions over the neighbors of atom $i$ as
\begin{equation}
\label{eq12b}
D^{(2)}_{ik}(\bm \eta^{(2)})  = \sum_{j} U^{(2)}_k(r_{ij},  \bm \eta^{(2)}), \quad 1 \le i \le N, 1 \le k \le K. 
\end{equation}
Owing to the rapid convergence of the KL expansion, only a small number of descriptors is needed to obtain accurate approximation. We therefore recommend to use no more than 10 two-body proper orthogonal descriptors. This procedure for creating two-body PODs is summarized in Figure \ref{fig:pod_procedure}, with more details on the implementation in Section 4.1. The logic behind the procedure in Figure \ref{fig:pod_procedure} also applies to three-body PODs, which we describe next.

\subsection{Three-body proper orthogonal descriptors}

In order to provide proper orthogonal descriptors for three-body interactions, we need to introduce a three-body parametrized potential. In particular, the three-body potential is defined as a product of radial and angular functions as follows
\begin{equation}
\label{eq9}
\begin{split}
V^{(3)}(r_{ij}, r_{ik}, \theta_{ijk}, \bm \eta^{(3)}, \bm \mu^{(3)})  =  \psi(r_{ij}, r^{(3)}_{\rm min}, r^{(3)}_{\rm max}, \alpha, \beta, \gamma, \kappa) f_{\rm c}(r_{ij}, r^{(3)}_{\rm min}, r^{(3)}_{\rm max}) \\
\psi(r_{ik}, r^{(3)}_{\rm min}, r^{(3)}_{\rm max}, \alpha, \beta, \gamma, \kappa) f_{\rm c}(r_{ik}, r^{(3)}_{\rm min}, r^{(3)}_{\rm max}) \\
\cos (\sigma \theta_{ijk} + \zeta) 
\end{split}
\end{equation}
where $\sigma$ is the periodic multiplicity, $\zeta$ is the equilibrium angle, $\bm \eta^{(3)} = (r^{(3)}_{\rm min}, r^{(3)}_{\rm max})$, and $\bm \mu^{(3)} = (\alpha, \beta, \gamma, \kappa, \sigma, \zeta)$.
The three-body potential (\ref{eq9}) provides an angular fingerprint of the atomic environment through the bond angles $\theta_{ijk}$ formed with each pair of neighbors $j$ and $k$.  Compared to the two-body potential (\ref{eq6}), the three-body potential (\ref{eq9}) has two extra parameters $(\sigma, \zeta)$ associated with the angular component. Note that $r_{\min}^{(3)}$ and $r_{\max}^{(3)}$ are the inner and outer cutoff distances for the three-body potential, respectively. 

Let $\bm \varrho = (\alpha, \beta, \gamma, \kappa)$. We assume that we are given $L^r$ parameter tuples $\bm \varrho_\ell, 1 \le \ell \le L^r$. We introduce the following set of  snapshots on $(r^{(3)}_{\min}, r^{(3)}_{\max})$:
\begin{equation}
\label{eq11q}
\zeta_\ell(r_{ij}, r^{(3)}_{\rm min}, r^{(3)}_{\rm max} ) =  \psi(r_{ij}, r^{(3)}_{\rm min}, r^{(3)}_{\rm max}, \bm \varrho_\ell) f_{\rm c}(r_{ij}, r^{(3)}_{\rm min},  r^{(3)}_{\rm max}), \quad 1 \le \ell \le L^r .
\end{equation}
We apply the Karhunen-Lo\`eve (KL) expansion to this set of snapshots to obtain orthogonal basis functions as follows
\begin{equation}
\label{eq12qq}
U^{r}_k(r_{ij}, r^{(3)}_{\rm min}, r^{(3)}_{\rm max} ) = \sum_{\ell = 1}^{L^r} a^r_{\ell k} \,  \zeta_\ell(r_{ij}, r^{(3)}_{\rm min}, r^{(3)}_{\rm max} ), \qquad k = 1, \ldots, K^r , 
\end{equation}
where the matrix components $a^r_{\ell k}$ consists of eigenvectors of the eigenvalue problem (\ref{eq3a:7}), and the number of basis functions $K^r$ consists of the top eigenvalues like with the two-body basis functions. Similarly, we  use the KL expansion to construct a set of orthogonal basis functions for the parametrized angular function $\cos (\sigma \theta_{ijk} + \zeta)$ defined on $(0, \pi)$. We denote the resulting angular basis functions as $U^{a}_n(\theta_{ijk}), 1 \le n \le K^a$, where $K^a$ is the number of angular basis functions.


The orthogonal basis functions for the parametrized potential (\ref{eq9}) are computed as follows
\begin{equation}
\label{eq14}
U^{(3)}_{mn}(r_{ij}, r_{ik}, \theta_{ijk}, \bm \eta^{(3)}) = U^{r}_m(r_{ij}, \bm \eta^{(3)}) U^{r}_m(r_{ik}, \bm \eta^{(3)}) U^{a}_n(\theta_{ijk}),
\end{equation}
for $1 \le m \le K^r, 1 \le n \le K^a$. Finally, the three-body proper orthogonal descriptors at each atom $i$ are obtained by summing (\ref{eq14}) over the neighbors $j$ and $k$ of atom $i$ as
\begin{equation}
\label{eq15}
D^{(3)}_{imn}(\bm \eta^{(3)})  = \sum_{j,k} U^{(3)}_{mn}(r_{ij}, r_{ik}, \theta_{ijk}, \bm \eta^{(3)}), \quad  1 \le m \le K^r, 1 \le n \le K^a.
\end{equation}
The number of three-body descriptors is $K^r K^a$. Using 50–100 three-body descriptors is sufficient in most cases. We therefore recommend to use no more than 100 three-body descriptors, because of the rapid convergence of the KL expansion like we saw with the two-body descriptors.

\subsection{Linear expansion of the descriptors}

All proper orthogonal descriptors are invariant with respect to translation and rotation of the atomic environment because they depend on the atomic distances and angles. Furthermore, because of the sum over all neighbors, they are invariant with respect to any permutation of chemically equivalent atoms in the environment. As a result, they can be used to define an expansion for the atomic energies that are invariant with respect to translation, rotation, and permutation. The most simple expansion is a linear combination of the descriptors leading to the following expression of the atomic energies 
\begin{equation}
\label{eq16}
E_{i}(\bm \eta) = c^{(1)} D^{(1)}_{i} + \sum_{m} c^{(2)}_m D^{(2)}_{im}(\bm \eta^{(2)})  + \sum_{mn} c^{(3)}_{mn} D^{(3)}_{imn}(\bm \eta^{(3)}), \quad 1 \le i \le N, 
\end{equation}
where $D^{(1)}_{i}, D^{(2)}_{im}, D^{(3)}_{imn}$ are the  descriptors associated with one-body, two-body  three-body parametrized potentials, respectively, and $c^{(1)}, c^{(2)}_m, c^{(3)}_{mn}$ are their respective expansion coefficients. In a more compact notation that implies summation over descriptor indices the atomic energies $E_i$ in (\ref{eq16}) can be written as
\begin{equation}
\label{eq18}
E_i(\bm \eta) = \sum_{m=1}^M c_m D_{im}(\bm \eta), \qquad 1 \le i \le N .
\end{equation}
Here $M$ is the total number of descriptors including one-body, two-body, and three-body descriptors. The potential energy is then obtained by summing local atomic energies $E_i$ for all atoms $i$ in the system
\begin{equation}
\label{eq17}
E(\bm \eta) = \sum_{i}^N E_{i}(\bm \eta) .
\end{equation}
 Because the descriptors are one-body, two-body, and three-body terms, the resulting potential energy (\ref{eq17}) is a three-body PES. 


In Section 4, we will describe how to implement an efficient calculation of energy and forces. We will also describe an fitting procedure to determine the expansion coefficients, as well as an optimization formulation to optimize the hyperparameters. 

\section{Relation to Other Descriptors}

In the section we will discuss some of the popular descriptors and the relation among them, as well as their connection to our method described in the previous section.

\subsection{Atom-centered symmetry functions and angular Fourier series}

The atom-centered symmetry function (ACSF) descriptors of Behler and Parrinello \cite{Behler2007,Behler2011,Behler2014} are derived from the radial function
\begin{equation}
\label{eq21}
G_i^{(2)} = \sum_j e^{-\eta (r_{ij}-r_s)^2} f_{\rm c}(r_{ij})   
\end{equation}
and the angular function
\begin{equation}
\label{eq22}
G_i^{(3)} = 2^{1 - \zeta} \sum_{j,k} (1 + \lambda \cos \theta_{ijk})^\zeta  \cdot  e^{-\eta (r_{ij}^2 + r_{ik}^2 + r_{jk}^2)} f_{\rm c}(r_{ij}) f_{\rm c}(r_{ik})  f_{\rm c}(r_{jk})      
\end{equation}
where the cut-off function is given by (\ref{eq7}). Different values of the parameters $\eta, r_s, \zeta, \lambda$ can be used to generate the ACSF descriptors.  Similar to the ACSF descriptors (\ref{eq22}), the angular Fourier series (AFS) descriptors are formed by using the Chebyshev polynomials of the angular component as $\mbox{AFS}_{inl}^{(3)} =g_n(r_{ij}) g_n(r_{ik}) \cos(l \theta_{ijk})$, where $g_n(r)$ are radial basis functions introduced in \cite{Bartok2013}.


Both ACSF and AFS are related to our method. However, because ACSF basis functions are generated directly from snapshots of the symmetry functions (\ref{eq21}) and (\ref{eq22}), they are not orthogonal. Moreover, both ACSF and ASF do not make use of the KL expansion to reduce the number of basis functions.

\subsection{Power spectrum and bispectrum descriptors}

The starting point for constructing the power spectrum and bispectrum descriptors is to consider the neighborhood density at location $\bm r$ around a central atom $i$: 
\begin{equation}
\label{s03d}
\rho_i(\bm r) =  \sum_{j}  \delta(\bm r - \bm r_{ij}),   
\end{equation}
where $\delta$ is the Dirac-delta function. Expanding the neighbor density  in terms of the spherical harmonics $Y_{lm}(\hat{\bm{r}})$ and radial basis functions $g_k(r)$ yields
\begin{equation}
\rho_i({\bm r}) = \sum_{k=0} \sum_{l=0} \sum_{m=-l}^l c_{iklm} g_k(r) Y_{lm}(\hat{\bm{r}}),
\end{equation}
where the expansion coefficients are given by the inner products of the neighborhood density with $g_k(r) Y_{lm}(\hat{\bm{r}})$ and expressed as 
\begin{equation}
\label{eq25}
 c_{iklm} = \sum_{j} g_k(r_{ij}) Y_{lm}(\hat{\bm{r}}_{ij}).
\end{equation}
The power spectrum and bispectrum descriptors are obtained as follows
\begin{equation}
\begin{split}
p_{i k_1k_2l} & = \sum_{m=-l}^l ({c}_{ik_1lm})^*  c_{ik_2lm} ,     \\
b_{i k_1 k_2 l_1l_2l_3} & = \sum_{m_1=-l_1}^{l_1} \sum_{m_2=-l_2}^{l_2} \sum_{m_3=-l_3}^{l_3}  ({c}_{ik_1l_1m_1})^* C^{l_1l_2l_3}_{m_1m_2m_3} c_{i k_2 l_2 m_2} c_{i k_2 l_3 m_3} ,
\end{split}
\end{equation}
where $C^{l_1l_2l_3}_{m_1m_2m_3}$ are the Clebsch-Gordan coefficients and the asterisk denotes the complex conjugation \cite{Bartok2013}. 

The power spectrum descriptors are three-body, while the bispectrum descriptors are four-body. An advantage of the power spectrum and bispectrum descriptors is that their evaluation scales linearly with the number of neighbors. However, they require computing a number of inner products and triple products. As the power spectrum descriptors can be expressed as products of polynomials of $\cos(\theta_{ijk})$ and radial basis functions, they are  related to the ACFS, AFS, and POD descriptors.

\subsection{Smooth overlap of atomic positions}

Instead of the Dirac-delta functions, the SOAP method \cite{Bartok2013} constructs the neighborhood density using Gaussians expanded in terms of spherical harmonic
functions as follows
\begin{equation}
\label{s03e}
\rho_i(\bm r) =  \sum_{j} \exp(- \alpha |\bm r - \bm r_{ij}|^2) =  \sum_{k l m}  c_{iklm} g_k(r) Y_{lm}(\hat{\bm{r}}) .
\end{equation}
The expansion coefficients are  inner products of the  neighborhood density with $g_k(r) Y_{lm}(\hat{\bm{r}}$) and can be computed by noting that
\begin{equation}
\label{s03f}
 \exp(- \alpha |\bm r - \bm r_{ij}|^2) =   4 \pi \exp(-\alpha (r^2 + r_{ij}^2)) \sum_{lm} I_l(2 \alpha r r_{ij}) Y_{lm}(\hat{\bm{r}}) Y_{lm}(\hat{\bm{r}}_{ij})
\end{equation}
where $I_l$ is a modified spherical Bessel function of the first kind. 


The SOAP descriptors are equivalent to using the  power or bispectrum descriptors together with Gaussian atomic neighbor density contributions and a dot product covariance kernel. The Gaussian width $\alpha$ controls  the smoothness of the similarity measure between atomic neighbor environments. The ability to controlling the smoothness of the similarity measure is an important advantage of SOAP over the  power or bispectrum descriptors. SOAP has been used for the construction of a variety of accurate and transferable
interatomic potentials \cite{Bartok2017,Bartok2018,Deringer2017,Dragoni2018}.

\subsection{Spectral neighbor analysis potential}

The SNAP descriptors \cite{Thompson2015} are constructed from hyperspherical harmonics on the unit 3-sphere. A 3-dimensional vector $\bm r$  is mapped onto the unit 3-sphere $S^3$ defined by three angles $(\phi,\theta,\omega)$ as follows 
\begin{equation}
\phi = \arctan (y/x), \quad \theta = \arccos(z/|\bm r|), \quad  \omega = \pi |\bm r|/ r_0 .
\end{equation}
In \cite{Bartok2013} it is recommended to use  $r_0 = \frac{4}{3} r_{\rm cut}$ to cover most of the sphere surface, while still excluding the region near the south pole. The Laplace–Beltrami operator on $S^3$ is defined as
\begin{equation}
\label{4dlap}
\Delta_{S^3} = \frac{1}{\sin^2 \omega} \frac{\partial }{\partial \omega} \sin^2 \omega \frac{\partial }{\partial \omega} + \frac{1}{\sin^2 \omega} \Delta_{S^2} ,
\end{equation}
where $\Delta_{S^2}$ is the Laplace–Beltrami operator on the unit 2-sphere $S^2$. The eigenfunctions of (\ref{4dlap}) are the hyperspherical harmonics
\begin{equation}
\label{4dlap1}
U_{jmm'}(\omega, \phi, \theta) = (\sin \omega)^m G_{j-m}^{m+1}(\cos \omega) Y_{mm'}(\theta,\phi)
\end{equation}
where $G_{j-l}^{l+1}$ are the Gegenbauer polynomials, and $Y_{mm'}$ are the usual spherical harmonics. These hyperspherical harmonics are also known as the Wigner matrices \cite{Thompson2015}.  

To make the contribution from atoms at $r = r_{\rm cut}$
vanish smoothly to zero, it is necessary to augment the atomic
neighbor density function (\ref{s03d}) with a cutoff function
\begin{equation}
\label{4dd}
\rho_i(\bm r) = \delta (\bm r) + \sum_{j}  f_{\rm c}(r _{i})  \delta (\bm r - \bm r_{ij}) .
\end{equation}
This atomic neighborhood density is expanded in terms of hyperspherical harmonics on the unit 3-sphere as
\begin{equation}
\rho_i(\bm r) = \sum_{j=0, 1/2,\ldots}^\infty \sum_{m=-j}^j \sum_{m'=-j}^j c_{ijmm'}  U_{jmm'}(\omega, \phi, \theta)
\end{equation}
where
\begin{equation}
 c_{ijmm'}  =  U_{jmm'}(0,0,0) + \sum_{j} f_{\rm c}(r _{ij})  U_{jmm'}(\omega_{ij}, \phi_{ij}, \theta_{ij}) .
\end{equation}
The coefficients  $c_{ijmm'}$ are not directly useful as descriptors, because they are not invariant under rotation of the reference  frame. However, the following bispectrum components are invariant with respect to permutation, translation and rotation
\begin{equation}
b_{ij_1j_2j}  = \sum_{m_1,m_1'=-j_1}^{j_1} \sum_{m_2,m_2'=-j_2}^{j_2} \sum_{m,m'=-j}^{j}  \left({c}_{ijmm'} \right)^* H\ts^{j m m'}?{j_1m_1m_1'}_{j_2m_2m_2'} \ c_{i j_1m_1m'_1} c_{i j_2 m_2m_2'} ,
\end{equation}
where the coupling constants are related to the Clebsch–Gordan coefficients as follows
\begin{equation}
  H\ts^{j m m'}?{j_1m_1m_1'}_{j_2m_2m_2'}  = C^{j j_1j_2}_{m m_1 m_2} C^{j j_1 j_2}_{m' m'_1 m'_2} .
\end{equation}
We see from (\ref{4dlap1}) that the hyperspherical harmonics can be expressed as a radial function time a spherical harmonic  $U_{jmm'}(\omega, \phi, \theta) = g_{k}(r) Y_{mm'}(\theta, \phi) $ with $g_{k}(r) = (\sin \omega)^m G_{j-m}^{m+1}(\cos \omega)$, where the index $k$ has a one-to-one correspondence to $(j,m)$. As a result, SNAP can be viewed as a variant of the bispectrum and SOAP descriptors. Recently, SNAP has been extended to quadratic SNAP \cite{Wood2018} and multi-element explicit SNAP \cite{Cusentino2020}.

\subsection{Atomic cluster expansion}


The atomic cluster expansion (ACE) is recently developed in \cite{Drautz2019,Drautz2020} as a hierarchical and complete descriptor of the local atomic environment due to its arbitrary number of body orders. This method extends the power and bispectrum construction to obtain a complete set of invariant descriptors. Similar to (\ref{eq25}), the expansion coefficients in the ACE method are given by
\begin{equation}
A_{inlm} = \sum_{j} R_{nl}(r_{ij}) Y_{lm} (\hat{\bm r}_{ij})
\end{equation}
where the radial basis functions $R_{nl}(r)$ are constructed from the Chebyshev polynomials of the first
kind and the cosine cut-off function. The ACE descriptors are then computed as
\begin{equation}
B^{(1)}_{in} = A_{in00},
\end{equation}
\begin{equation}
B^{(2)}_{in_1n_2l} = \sum_{m=-l}^l (-1)^m A_{i n_1l m} A_{i n_2l -m},
\end{equation}
\begin{equation}
B^{(3)}_{\substack{in_1n_2n_3\\ l_1 l_2 l_3}} = \sum_{m_1=-l_1}^{l_1} \sum_{m_2=-l_2}^{l_2} \sum_{m_3=-l_3}^{l_3} W_{m_1m_2m_3}^{l_1l_2l_3} A_{i n_1l_1 m_1} A_{i n_2l_2 m_2} A_{i n_3l_3 m_3},
\end{equation}
\begin{equation}
B^{(4)}_{\substack{in_1n_2n_3n_4 \\ l_1 l_2 l_3l_4}} = \sum_{m_1m_2m_3m_4}  W_{m_1m_2m_3m_4}^{l_1l_2l_3l_4} A_{i n_1l_1 m_1} A_{i n_2l_2 m_2} A_{i n_3l_3 m_3} A_{i n_4l_4 m_4}, 
\end{equation}
which are in that order the two-body, three-body, four-body, and five-body descriptors, respectively. The coupling constants $W_{m_1m_2m_3}^{l_1l_2l_3}$ and $W_{m_1m_2m_3m_4}^{l_1l_2l_3l_4}$ are calculated from the Clebsch–Gordan coefficients. The paper \cite{Drautz2019} describes a systematic approach to defining many-body ACE descriptors. 

The ACE method has a number of attractive features. First, the method encompasses several well-known descriptors since ACSF, SOAP, power spectrum, bispectrum, and SNAP can be expanded using the ACE descriptors. Second, the method is complete in the sense that it is possible to converge cluster functional to arbitrary accuracy. And third, the ACE descriptors scale linearly with the number of neighbors irrespective of the order of the expansion. The ACE method has been analyzed in \cite{DUSSON2022110946} and applied to many different problems \cite{Kovacs2021,Lysogorskiy2021,Bochkarev2022}.

\section{Implementation}

In this section, we describe the implementation of the proper orthogonal descriptors and the potential fitting procedure.

\subsection{Precomputation of the eigenvectors and eigenvalues}

To compute the eigenvectors in (\ref{eq12}), we need to define the parameter space and choose $L$ parameter points $\bm \mu^{(2)}_\ell = (\alpha_\ell, \beta_\ell, \gamma_\ell, \kappa_\ell), 1 \le \ell \le L$. The parameters $\alpha \in [1, P_\alpha]$ and $\gamma \in [1, P_\gamma]$ are integers, where $P_\alpha$ and $P_\gamma$ are the highest degrees for $\alpha$ and $\gamma$, respectively. We recommend to uniformly choose $P_\beta$ different values of $\beta$ in the interval $[\beta_{\min}, \beta_{\max}]$, where $\beta_{\min} = 0$ and $\beta_{\max} = 5$. The parameter $\kappa$ can be set either 0 or 1. Hence, the total number of parameter points is $L = P_\alpha P_\beta + P_\gamma$. Although  $P_\alpha$, $P_\beta$, $P_\gamma$ can be chosen conservatively large, we recommend to choose $P_\alpha \le 8$, $P_\beta \le 4$, $P_\gamma \le 12$. In many cases, $P_\alpha = 6$, $P_\beta = 3$, $P_\gamma = 8$ are adequate. Once the parameter points are chosen, we compute the inner products of pairs of snapshots, $\{\xi_\ell(r, r_{\min}, r_{\max})\}_{\ell=1}^L$ to form the eigenvalue problem $\ref{eq3a:7}$. At this point, $r_{\rm min}$ and $r_{\rm max}$ are assumed to be fixed. As discussed later in section 4.4, both $r_{\rm min}$ and $r_{\rm max}$ are found by solving an outer optimization problem.

The eigenvectors in (\ref{eq12qq}) can be computed using the same procedure as described above. Finally, we compute the eigenvectors for the angular component $\cos(\sigma \theta + \zeta)$ on the domain $\theta \in (0, \pi)$ in a similar manner. Typically, we choose $\sigma$ to be integers in $[1, P_\sigma]$ while fixing $\zeta = 0$, where $P_\sigma$ is the highest degree for $\sigma$. We recommend to choose $P_\sigma \le 10$ in most cases.

\subsection{Calculation of energy and forces}

The radial basis functions in (\ref{eq12}) are first calculated. Owing to the exponential convergence of the KL expansion, the number of basis functions $K$ is considerably smaller than the number of snapshots $L$. Therefore, the two-body descriptors (\ref{eq12b}) can be computed by summing the radial basis functions over neighbors of every atom $i$. Next, we compute the radial and angular basis functions, which are then used to compute the three-body basis functions (\ref{eq14}). Finally, the three-body descriptors (\ref{eq15}) are obtained by summing the three-body basis functions over neighbors $j$ and $k$ of every atom $i$.


We can now calculate the local atomic energies for the linear expansion of the descriptors (\ref{eq18}). The potential energy is then obtained by summing the atomic energies as
\begin{equation}
E(\bm \eta) =   \sum_{m=1}^M c_m \sum_{i=1}^N   D_{im}(\bm \eta) =  \sum_{m=1}^M c_m d_{m}(\bm \eta) ,   
\end{equation}
where $d_{m}(\bm \eta) = \sum_{i=1}^N D_{im}(\bm \eta)$ are the global descriptors. To calculate forces, we note that 
\begin{equation}
\bm F(\bm \eta) = -\nabla E(\bm \eta) =  - \sum_{m=1}^M c_m \sum_{i=1}^N  \nabla D_{im}(\bm \eta) = - \sum_{m=1}^M c_m \nabla d_{m}(\bm \eta) ,   
\end{equation}
where $\nabla d_{m}(\bm \eta)$ are derivatives of the global descriptors with respect to atom positions. For the two-body global descriptors, their derivatives are computed from the derivatives of the radial basis functions (\ref{eq12}) with respect to $r_{ij}$ for all neighbor pairs $i,j$. For the three-body global descriptors, their derivatives are computed from the derivatives of the radial basis functions with respect to $r_{ij}$ and $r_{ik}$, and the derivatives of the angular basis functions with respect to $\theta_{ijk}$.


\subsection{Weighted least-squares regression}

With the availability of high-performance computers and highly optimized DFT codes, it is not difficult to generate DFT data for hundreds or thousands of configurations of atoms. Let $J_{\rm train}$ be the total number of training configurations. Let $\{E^{\rm DFT}_j\}_{j=1}^{J_{\rm{train}}}$ and $\{\bm F^{\rm DFT}_j\}_{j=1}^{J_{\rm{train}}}$ be the DFT energies and forces for $J_{\rm train}$ configurations. Next, we calculate the global descriptors and their derivatives for all training  configurations. We then form the matrix $\bm A \in \mathbb{R}^{J_{\rm train} \times M}$, where the $j$th row of $\bm A$ consists of the global descriptors for the $j$th configuration. Similarly, we form the matrix $\bm B \in \mathbb{R}^{N_J \times M}$ by stacking the derivatives of the global descriptors for all training configurations from top to bottom, where  $N_J = 3\sum_{j=1}^{J_{\rm train}} N_j$ and $N_j$ is the number of atoms for the $j$th configuration. Note that both $\bm A$ and $\bm B$ depends on $\bm \eta$.

When the hyperparameters $\bm \eta$ are known, the coefficient vector $\bm c$ of the linear expansion (\ref{eq18}) is found by solving the following least-squares problem  
\begin{equation}
\label{eq45}
\min_{\bm c \in \mathbb{R}^{M}}    \beta^E \|\bm A(\bm \eta) \bm c - \bm E^{\rm DFT} \|^2 + \beta^F \|\bm B(\bm \eta) \bm c - \bm F^{\rm DFT} \|^2, 
\end{equation}
where $\beta^E$ is the energy weight and $\beta^F$ is the force weight. Here $\bm E^{\rm DFT}$ is a vector of $J_{\rm train}$ DFT energies and $\bm F^{\rm DFT}$ is a vector of $N_J$ DFT forces. In our implementation, we fix $\beta^F = 1$ and let $\beta^E$ be the only scalar weight to be dependent on specific problems. 

\subsection{Optimization of hyperparameters}

The hyperparameters contain inner and outer cut-off distances for two-body and three-body interactions. They can have a critical effect on the prediction performance of the resulting potential. We observe through our numerical experiences that longer cut-off distances do not necessarily translate to better performance. Furthermore, setting them to some heuristic values may lead to an interatomic potential that can be considerably less accurate than the potential obtained by optimizing the cut-off distances. Therefore, it is crucial to optimize the hyperparameters.    

Let $\Omega^{\bm \eta}$ be a domain in which the hyperparameters reside. We consider solving the following optimization problem
\begin{equation}
\label{eq46}
\min_{\bm \eta \in \Omega^{\bm \eta}}    w \|\bm A(\bm \eta) \bm c(\bm \eta) - \bm E^{\rm DFT} \|_{\rm MAE} + (1-w) \|\bm B(\bm \eta) \bm c(\bm \eta) - \bm F^{\rm DFT} \|_{\rm MAE}
\end{equation}
where $\bm c(\bm \eta)$ is the solution of the least-squares problem (\ref{eq45}), $w \in [0,1]$ is a given scalar weight, and $\|\cdot \|_{\rm MAE}$ is defined as
\begin{equation}
\|\bm e\|_{\rm MAE} = \frac1n \sum |e_i| .
\end{equation}
The problem (\ref{eq46}) is an outer optimization problem, while (\ref{eq45}) is an inner optimization problem.

Instead of directly solving the problem (\ref{eq46}), we solve a surrogate problem. To this end, we approximate the loss function in (\ref{eq46}) using polynomial tensor-product interpolation. In particular, we divide the domain $\Omega^{\bm \eta}$ into $Q$ grid points $\{\bm \eta_q\}_{q=1}^Q$ and compute the loss function at those grid points. This requires us to solve the least-squares problem (\ref{eq46}) $Q$ times for $\bm \eta = \bm \eta_q, q = 1, \ldots, Q$. We then fit the loss values at the grid points to a tensor-product multivariate polynomial of certain degree to construct a surrogate model of the loss function. Finally, the gradient descent algorithm is used to minimize the surrogate model to obtain the hyperparameters. In order to be able to reach the global minimum, we start our gradient descent algorithm with a large number of initial guesses randomly  sampled in $\Omega^{\bm \eta}$. This is necessary because the loss function (\ref{eq46}) may have multiple local minima.

\section{Numerical Experiments}

 In this section, we assess the performance of the ACE, SNAP, as well as POD descriptors for fitting interatomic potentials on previously published data sets for many different kinds of elements. In all cases, the weighted least-square regression (\ref{eq45}) is solved with $\beta^E = 100$ and $\beta^F = 1$ for the various descriptors considered here. For the ACE and POD methods, both the inner and outer cut-off distances are optimized by solving a polynomial interpolation surrogate model of the problem (\ref{eq46}) with $w = 0.8$ and 100 randomly sampled initial guesses. For the SNAP method, however, only the outer cut-off radius is optimized. We shall evaluate these methods on the basis of mean absolute errors of the predicted energy and forces versus the number of descriptors. To facilitate the reproduction of our work, the code and the results presented in this section are made available as  an open-source software on Github (https://github.com/
cesmix-mit/mlp). 


\subsection{Tantalum element}

The Ta data set contains a wide range of configurations to adequately sample the important regions of the potential energy surface \cite{Thompson2015}. The data set includes 12 different groups such as surfaces, bulk structures, defects, elastics for BCC, FCC, and A15 crystal structures, and high temperature liquid. Displaced configurations were constructed by randomly displacing atoms from their equilibrium lattice sites in supercells of the A15, BCC, and FCC crystal structures. Elastic configurations were constructed by applying random strains to primitive cells of the BCC and FCC crystals. The configurations of type “GSF” consist of both relaxed and unrelaxed generalized stacking faults along the [110] and [112] crystallographic directions.  Liquid configurations were taken from a high-temperature quantum molecular dynamics simulations of molten tantalum. Surface configurations consisted of relaxed and unrelaxed [100], [110], [111], and [112] BCC surfaces. Volume configurations consisted of primitive cells of A15, BCC, and FCC crystal structures, compressed or expanded isotropically over a wide range of densities. The database was used to create a SNAP potential \cite{Thompson2015} which successfully describes a wide range of properties such as energetic properties of solid tantalum phases, the size and shape of the Peierls barrier for screw dislocation motion in BCC tantalum, as well as both the structure of molten tantalum and its melting point. 

Herein  the same database is used to fit ACE, POD, and SNAP potentials. We associate a weight of 100.0 with energy, 1.0 with force, and 0.0 with stress for all configurations in the database. This is different from the SNAP potential presented in \cite{Thompson2015}, that specified different weights for each type of configuration. Furthermore, while the Ziegler–Biersack–Littmark (ZBL) empirical potential was used as the reference potential for the SNAP potential in \cite{Thompson2015}, no reference potential is used for ACE, POD, and SNAP potentials in this paper. The SNAP potential constructed in this paper has considerably smaller energy errors than the one in \cite{Thompson2015}, while having similar force errors. In particular, the MAE energy error of our SNAP potential can be close to 1 meV/Atom, while that of the SNAP potential in \cite{Thompson2015} is about 100 meV/atom.

\begin{table}[htbp]
\centering
	\begin{tabular}{|ccc|ccc|cc|}
		\cline{1-8}
		 \multicolumn{3}{|c|}{\textbf{ACE}} & \multicolumn{3}{c|}{\textbf{POD}} & 
		 \multicolumn{2}{c|}{\textbf{SNAP}}\\
		\hline
		$M$ & $r_{\min}$ & $r_{\max}$ & $M$ & $r_{\min}$ & $r_{\max}$ & $M$ & $r_{\max}$ \\
		\hline
		6 & 1.29 & 4.46 & 6 & 1.04 & 4.65 & 6 & 5.00 \\
		16 & 0.92 & 4.42 & 16 & 1.42 & 4.49 & 15 & 4.38\\
		32 & 1.19 & 4.53 & 32 & 1.14 & 5.00 & 31 & 4.33\\
		57 & 1.48 & 4.50 & 57 & 0.90 & 4.68 & 56 & 4.61\\
		93 & 0.99 & 4.43 & 92 & 0.90 & 4.75 & 92 & 4.64\\
		142 & 0.98 & 5.00 & 143 & 0.94 & 4.84 & 141 &4.60\\
		\hline
	\end{tabular}
	\caption{Optimized inner and outer cut-off distances of the ACE, POD, and SNAP potentials as a function of the number of descriptors for the Ta element.} 
	\label{tab1}
\end{table}

\begin{figure}[htbp]
	\centering
	\begin{subfigure}[b]{0.49\textwidth}
		\centering
		\includegraphics[width=\textwidth]{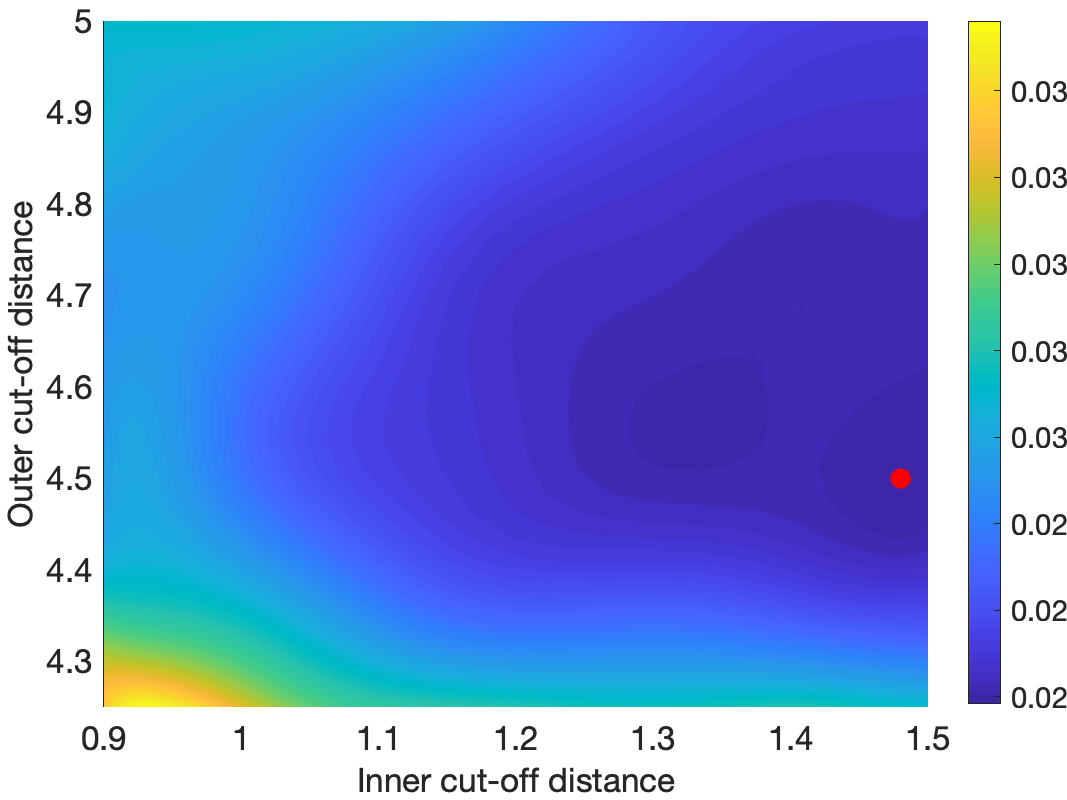}
		\caption{ACE potential with $M=57$ descriptors.}
	\end{subfigure}
	\hfill
	\begin{subfigure}[b]{0.49\textwidth}
		\centering
		\includegraphics[width=\textwidth]{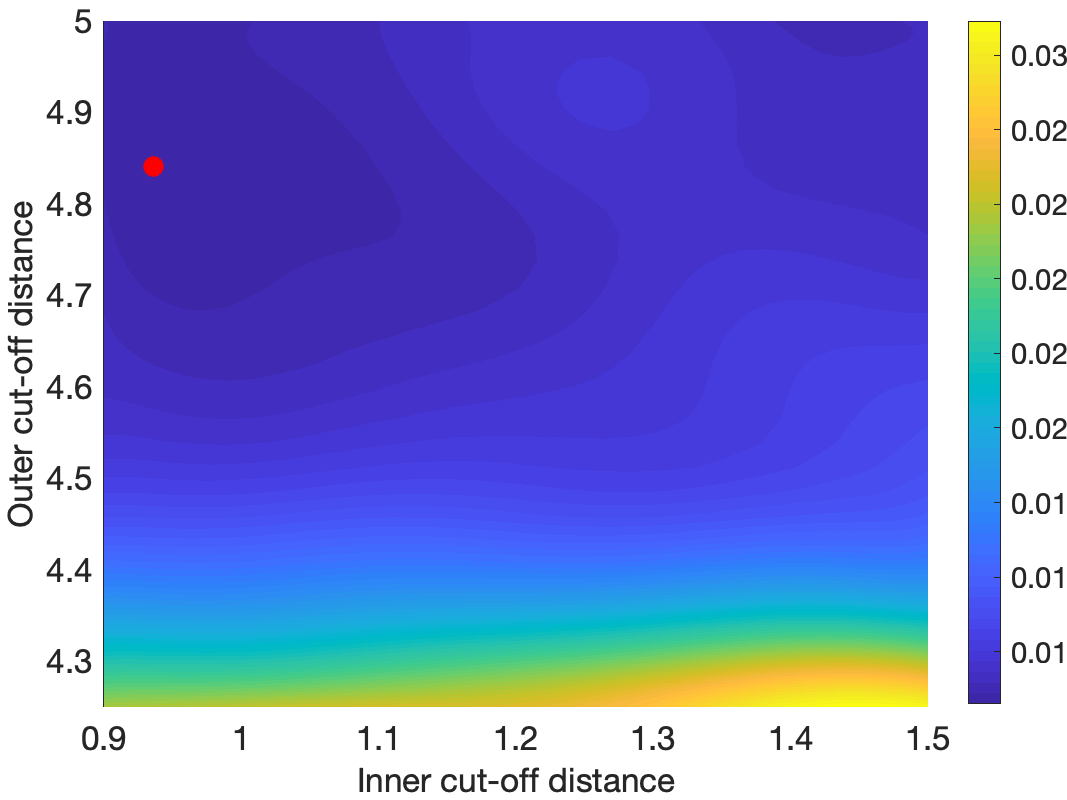}
		\caption{POD potential with $M=143$ descriptors.}
	\end{subfigure}
	\caption{Surrogate loss function of ACE and POD potentials as a function of inner and outer cutoff distances for the Ta element. The red circles indicate the optimal value of the cutoff distances for which the surrogate loss function is minimum}
	\label{fig1}
\end{figure}

Table \ref{tab1} provides the optimized cutoff distances of the ACE, POD, and SNAP potentials for several different values of the number of descriptors. The optimized cutoff distances for different potentials appear quite similar. In all cases, the optimized cutoff distances fall between the second nearest neighbor (2NN) and 3NN distances of Ta element. These observations are consistent with the previous SNAP potential \cite{Thompson2015} for Ta element, which has the cutoff radius of 4.67 \AA. While the optimized cutoff distances vary with the number of descriptors, the variation appears not monotonic.  Figure \ref{fig1} shows the polynomial interpolation surrogates of the loss function for ACE and POD potentials. We see that in both cases the surrogate loss functions have multiple local minima. Nonetheless, the surrogate loss functions are smooth. Using 100 randomly sampled initial guesses in the gradient descent algorithm, we can find the global minimum in both cases, which are indicated by the red circles in Figure \ref{fig1}. The results also show  significant influence  of the cutoff distances on the value of the loss function and thus on the accuracy of the resulting potential. Therefore, it is crucial to rigorously optimize these hyperparameters since heuristic means of choosing the hyperparameters may lead to suboptimal performance.

\begin{figure}[h]
	\centering
	\begin{subfigure}[b]{0.49\textwidth}
		\centering
		\includegraphics[width=\textwidth]{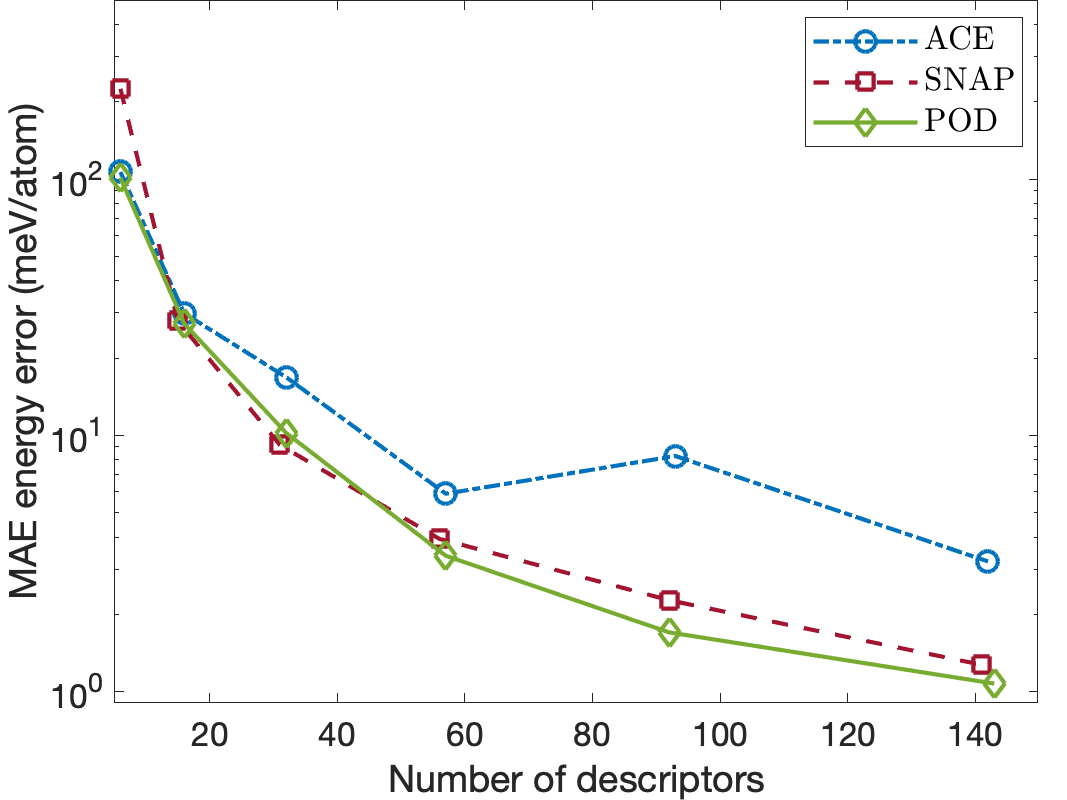}
		\caption{Mean absolute error in energy.}
	\end{subfigure}
	\hfill
	\begin{subfigure}[b]{0.49\textwidth}
		\centering
		\includegraphics[width=\textwidth]{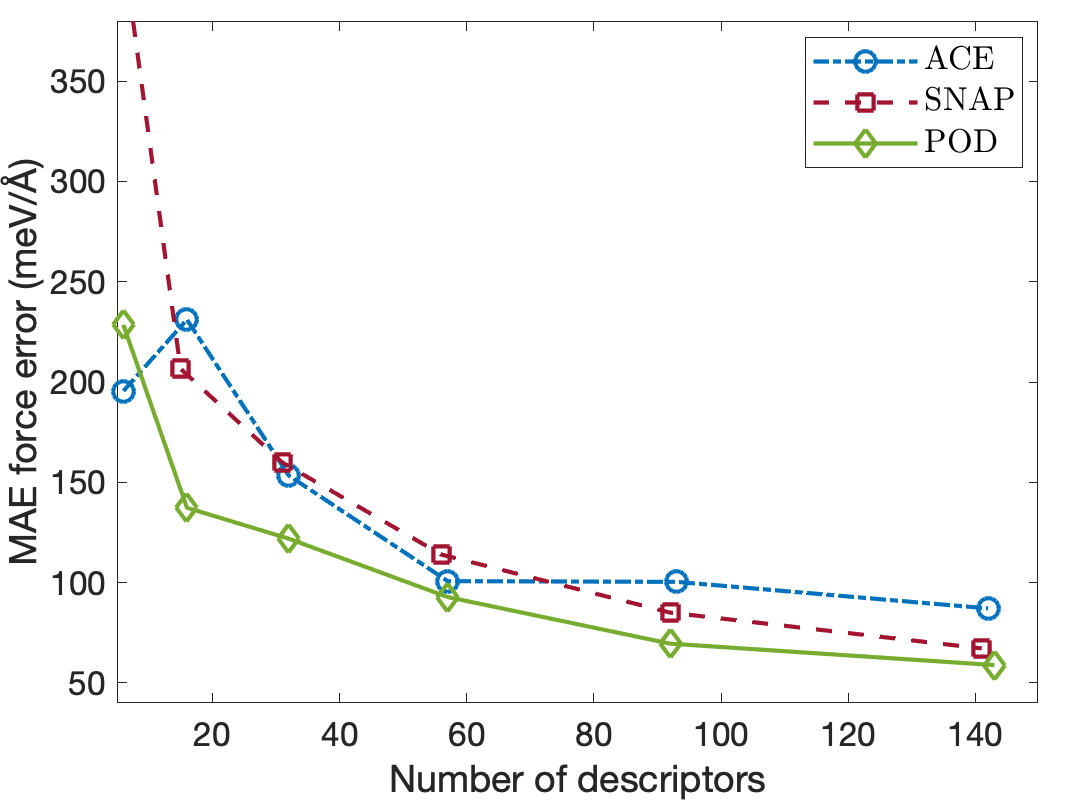}
		\caption{Mean absolute error in forces.}
	\end{subfigure}
	\caption{Energy errors in log scale and force errors in linear scale for ACE, SNAP, and POD potentials as a function of the number of descriptors for the Ta element.}
	\label{fig2}
\end{figure}
 
Figure \ref{fig2} depicts training errors in energies and forces predicted using ACE, SNAP, POD potentials for different values of the number of the descriptors. The same results are also tabulated in Table \ref{tab2} for quantitative comparison of the  potentials. We observe fast convergence of the energy and force errors as the number of descriptors increases from 16 to 60. However, the convergence becomes slower as the number of descriptors increases above 60. Both POD and SNAP have very similar convergence of the  energy error and yield faster convergence of the energy error than ACE. The convergence of the force error is faster for POD than ACE and SNAP, as  POD has smaller force errors than ACE and SNAP in most cases. The MAE of the energy reaches about 1.07 meV/atom for POD with $M = 143$, 1.27 meV/atom for SNAP with $M = 141$, and 3.21 meV/atom for ACE with $M = 142$.

\begin{table}[htbp]
\centering
\small
	\begin{tabular}{|ccc|ccc|ccc|}
		\cline{1-9}
		 \multicolumn{3}{|c|}{\textbf{ACE}} & \multicolumn{3}{c|}{\textbf{SNAP}} & 
		 \multicolumn{3}{c|}{\textbf{POD}}\\
		\hline
		$M$ & MAE(e) & MAE(f) & $M$ & MAE(e) & MAE(f) & $M$ & MAE(e) & MAE(f) \\
		\hline
6 & 106.52 & \textbf{195.51} & 6 & 225.13 & 415.27 & 6 & \textbf{101.14} & {228.50} \\
16 & 29.91 & 231.13 & 15 & 27.91 & 206.63 & 16 & \textbf{27.26} & \textbf{137.34}\\
32 & 16.83 & 153.45 & 31 & \textbf{9.16} & 159.89 & 32 & 10.19 &\textbf{121.93} \\
57 & 5.91 & 100.75 & 56 & 3.94 & 114.01 & 57 & \textbf{3.38} &\textbf{92.83} \\
93 & 8.29 & 100.39 & 92 & 2.27 & 85.06 & 92 & \textbf{1.70} & \textbf{69.59}\\
142 & 3.21 & 87.16 & 141 & 1.27 & 67.21 & 143 &\textbf{1.07} & \textbf{58.91}\\
		\hline
	\end{tabular}
	\caption{Training errors in  energies and forces for ACE, SNAP, and POD potentials as a function of the number of descriptors for the Ta element. The units of the mean absolute errors of the energy and force are meV/atom and meV/\AA, respectively.} 
	\label{tab2}
\end{table}

\begin{figure}[htbp]
\centering
	\begin{subfigure}[b]{0.49\textwidth}
		\centering
		\includegraphics[width=\textwidth]{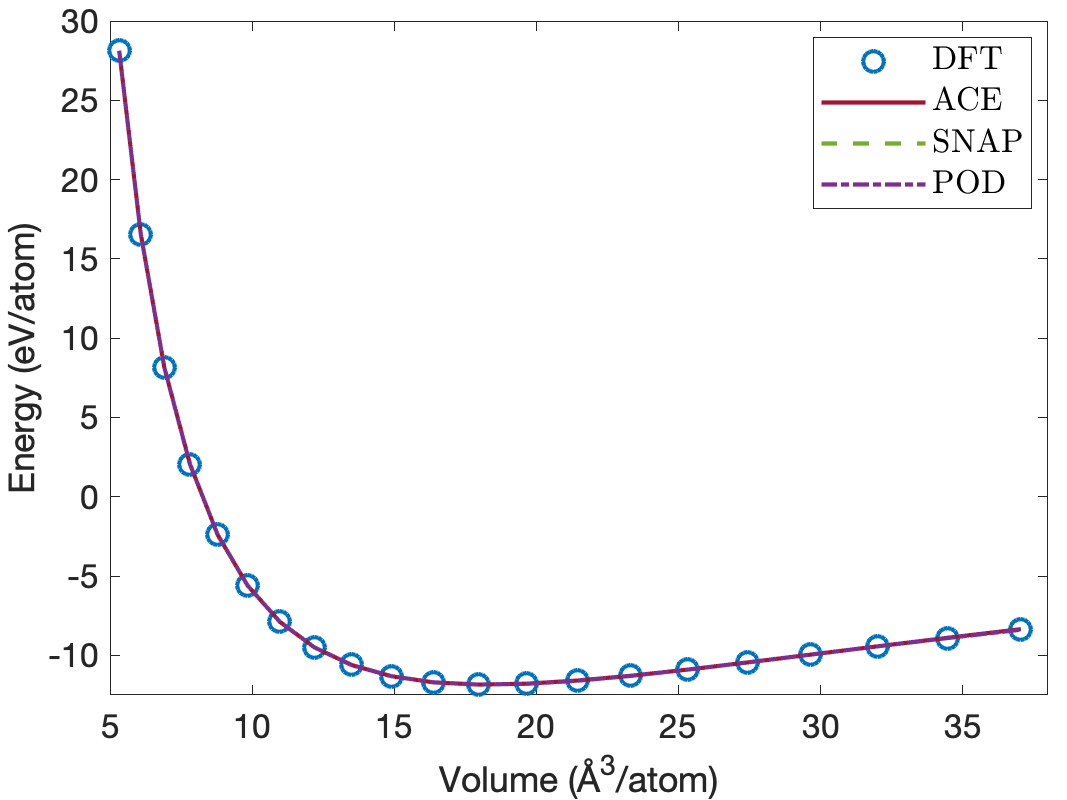}
		\caption{BCC.}
	\end{subfigure}
	\hfill
	\begin{subfigure}[b]{0.49\textwidth}
		\centering
		\includegraphics[width=\textwidth]{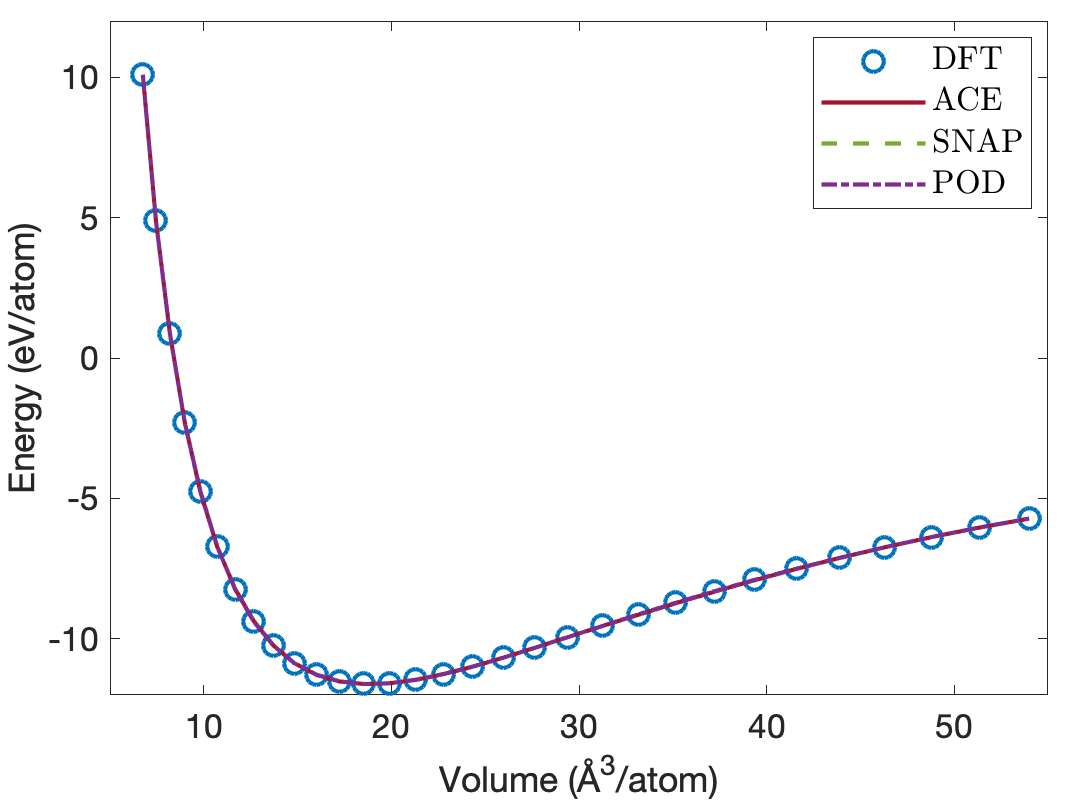}
		\caption{FCC.}
	\end{subfigure}
\caption{Energy per atom versus volume per atom for BCC and FCC crystal structures of the Ta element computed using DFT and ACE, SNAP,  POD potentials with the largest number of descriptors.}
\label{Ta:energy}	
\end{figure}

 Table \ref{tab2c} provides the training errors in energy and forces for each of the 12 groups. The Surface group has the the highest mean absolute error in energy, while the Liquid group has the highest mean absolute error in forces. The liquid structures depend strongly on the repulsive interactions that occur when two atoms approach each other. Consequently, it is more difficult for potential models to predict forces of the liquid structures since  the liquid configurations are very different from those of the equilibrium solid crystals.  It is also more difficult to predict energies of surface configurations because the surfaces of BBC crystals tend to be rather open with surface atoms exhibiting rather low coordination numbers. The MAE force errors for Volume A15, Volume BCC, and Volume FCC are zero, because all of the structures in Volume A15, Volume BCC, and Volume FCC are at equilibrium states and thus have zero atomic forces. Figure \ref{Ta:energy} plots the energy per atom computed with DFT, ACE, SNAP, and POD as a function of volume per atom for the BCC and FCC crystals. We see that the energies predicted using ACE, SNAP, POD potentials are extremely accurate.

\begin{table}[htbp]
\centering
\small
	\begin{tabular}{|c|cc|cc|cc|}
		\cline{1-7}
		Training  & \multicolumn{2}{|c|}{\textbf{ACE}} & \multicolumn{2}{c|}{\textbf{SNAP}} & 
		 \multicolumn{2}{c|}{\textbf{POD}}\\
		\cline{2-7}
		group & MAE(e) & MAE(f) & MAE(e) & MAE(f) & MAE(e) & MAE(f) \\
		\hline
Displaced A15 & 1.542 & 99.06 & 0.826 & 76.16 & 0.432 & 57.63 \\ 
 Displaced BCC & 9.211 & 148.4 & 2.365 & 104.9 & 0.879 & 100.4 \\ 
 Displaced FCC & 4.509 & 98.47 & 0.261 & 49.66 & 0.708 & 40.05 \\ 
 Elastic BCC & 0.600 & 47.19 & 0.243 & 32.80 & 0.219 & 42.76 \\ 
 Elastic FCC & 0.325 & 39.78 & 0.280 & 36.84 & 0.350 & 41.14 \\ 
 GSF 110 & 4.988 & 32.33 & 2.252 & 23.33 & 1.870 & 23.60 \\ 
 GSF 112 & 8.051 & 74.08 & 2.915 & 45.56 & 2.882 & 45.57 \\ 
 Liquid & 8.940 & 312.2 & 3.558 & 304.3 & 1.113 & 239.6 \\ 
 Surface & 12.39 & 48.30 & 9.713 & 53.35 & 10.68 & 42.66 \\ 
 Volume A15 & 4.718 & 0.000 & 2.689 & 0.000 & 2.113 & 0.000 \\ 
 Volume BCC & 7.676 & 0.000 & 2.336 & 0.000 & 1.087 & 0.000 \\ 
 Volume FCC & 7.539 & 0.000 & 1.775 & 0.000 & 1.444 & 0.000 \\ 
		\hline
	\end{tabular}
	\caption{Energy and force errors obtained using ACE, SNAP, POD potentials with the largest number of descriptors for different training groups for the Ta element. The units of the mean absolute errors of the energy and force are meV/atom and meV/\AA, respectively.} 
	\label{tab2c}
\end{table}

\subsection{Li, Mo, Ni, Cu, Si and Ge elements}

In the recent paper \cite{Zuo2020}, a diverse DFT data set was generated for Li, Mo, Ni, Cu, Si and Ge elements to evaluate and compare the performance of machine learning potentials based on four different descriptors.
These elements were chosen to represent a variety of chemistries (main group metal, transition metal and semiconductor), crystal structures (bcc, fcc, and diamond) and bonding types (metallic and covalent). For each element, a set of configurations with diverse coverage of atomic local environment space were generated. For each element, the full data set were divided into a training set and a test set in order to assess the ML potentials on both training and test configurations. We refer to \cite{Zuo2020} for additional details about the DFT data set, as well as insightful results on the performance and cost assessment of the various ML potentials studied therein. 

\begin{figure}[htbp]
\centering
\includegraphics[width=0.98\textwidth]{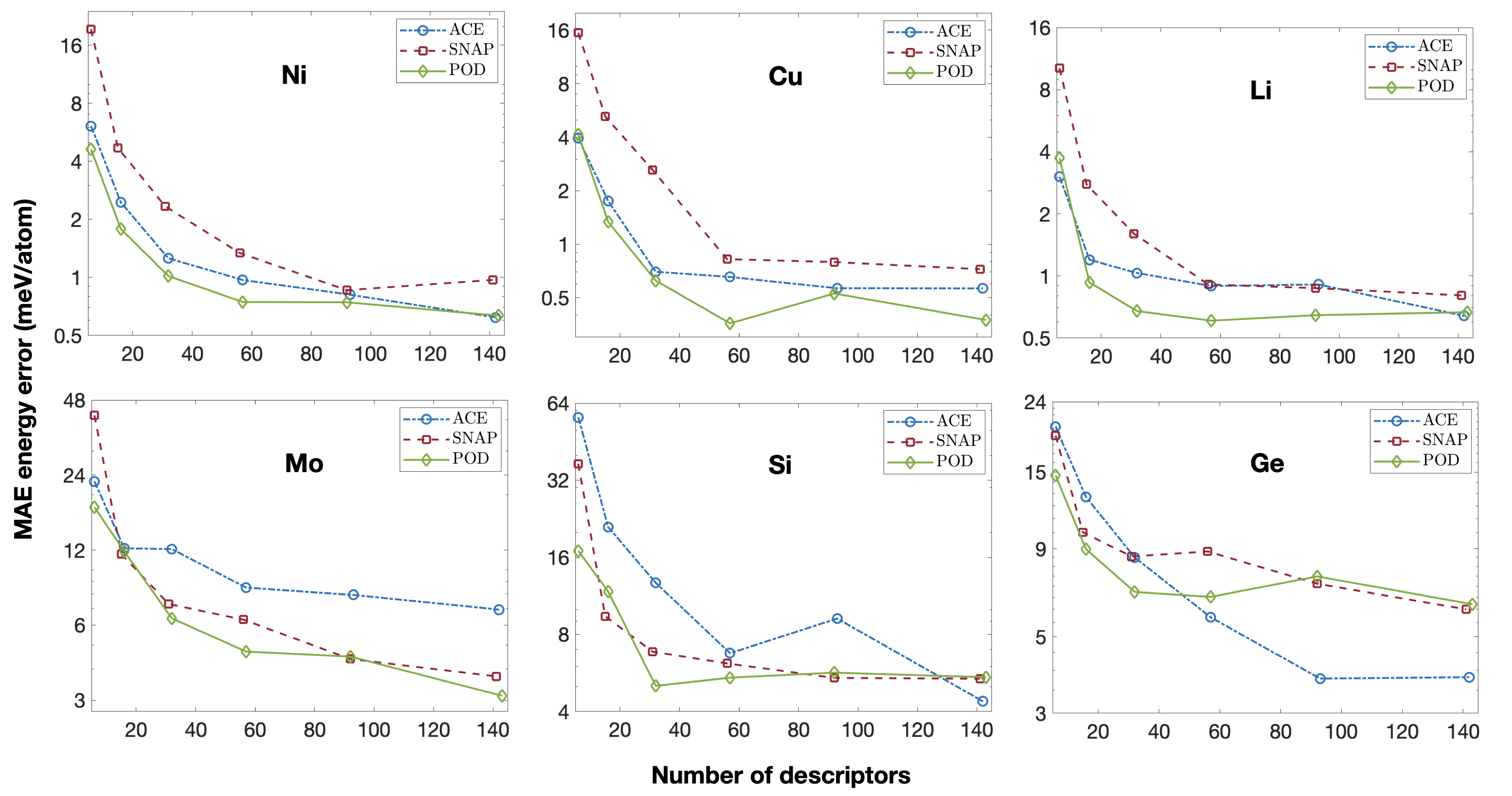}
\caption{Test errors in energies predicted using ACE, SNAP, and POD potentials as a function of the number of descriptors for Ni, Cu, Li, Mo, Si, Ge elements.}
\label{fig5}	
\end{figure}

We use ACE, SNAP, and POD descriptors to fit interatomic potentials for Li, Mo, Ni, Cu, Si and Ge elements. The energy errors on the test sets as a function of the number of descriptors are shown in Figure \ref{fig5}. We observe that the energy errors drop quickly when the number of descriptors increases up to 60. When the number of descriptors increases above 60, the convergence becomes slower. These observations are similar to those of the Ta element presented earlier. To quantitatively compare the descriptors, we present in Table \ref{tab4} the energy errors for three different values of the number of descriptors for both the training and test sets. The training errors are comparable to the testing errors across different descriptors, indicating no over-fitting for the potentials. The POD potentials generally have lower MAEs in energies than ACE and SNAP potentials for Ni, Cu, Li, Mo elements. The ACE potentials generally have lower energy errors than the SNAP and POD potentials for Ge element. All the potentials show extremely good performance for Ni, Cu, and Li elements, reaching about 1 meV/atom or less for the MAE of the energy. They have about 5 meV/atom in the MAE of the energy for Mo, Si, Ge elements, which are generally below the limits of DFT error.    

\begin{table}[htbp]
\centering
\small
	\begin{tabular}{|c|c| c c c c c c|}
		\cline{1-8}
		 Descriptors  & Data set & Ni & Cu & Li & Mo & Si & Ge \\
		\hline
		\hline
ACE  & Training & 2.374 & 1.718 & 1.198 & 13.52 & 18.28 & 11.81 \\
$M=16$ & Test & 2.453 & 1.754 & 1.199 & 12.19 & 21.02 & 12.71 \\
 \hline
SNAP & Training &  3.903 & 3.633 & 2.107 & 10.37 & 10.92 & 10.23 \\
$M = 15$ & Test & 4.704 & 5.234 & 2.783 & 11.57 & 9.428 & 10.03 \\
\hline 
POD & Training & 1.482 & 1.143 & 0.754 & 11.16 & 12.27 & 8.745 \\
 $M=16$ & Test & 1.784 & 1.338 & 0.931 & 11.87 & 11.77 & 8.972 \\
\hline		
\hline
ACE  & Training & 0.970 & 0.779 & 0.721 & 8.626 & 8.953 & 6.819 \\
$M=57$ & Test & 0.970 & 0.654 & 0.894 & 8.483 & 6.761 & 5.693 \\
 \hline
SNAP & Training & 1.128 & 1.001 & 0.818 & 5.510 & 4.657 & 7.916 \\
$M = 56$ & Test & 1.339 & 0.824 & 0.909 & 6.314 & 6.170 & 8.835 \\
\hline 
POD & Training & 0.699 & 0.427 & 0.601 & 4.110 & 5.948 & 6.759 \\
 $M=57$ & Test & 0.746 & 0.359 & 0.607 & 4.699 & 5.425 & 6.526 \\
 		\hline
 		\hline 
 ACE  & Training & 0.663 & 0.614 & 0.553 & 6.411 & 6.291 & 4.755 \\
$M=142$ & Test &  0.620 & 0.564 & 0.640 & 6.924 & 4.385 & 3.817 \\ 
 \hline
SNAP & Training & 0.916 & 0.835 & 0.657 & 3.747 & 3.973 & 6.586 \\
$M = 141$ & Test & 0.971 & 0.724 & 0.805 & 3.737 & 5.374 & 6.002 \\
\hline 
POD & Training & 0.490 & 0.408 & 0.579 & 3.085 & 5.380 & 6.094 \\
 $M=143$ & Test & 0.634 & 0.375 & 0.668 & 3.124 & 5.445 & 6.213 \\
 		\hline
	\end{tabular}
	\caption{Mean absolute errors in energies (meV/atom) predicted using the ACE, SNAP, POD methods for the training and test sets of Ni, Cu, Li, Mo, Si, Ge elements.} 
	\label{tab4}
\end{table}

\begin{figure}[h]
\centering
\includegraphics[width=0.99\textwidth]{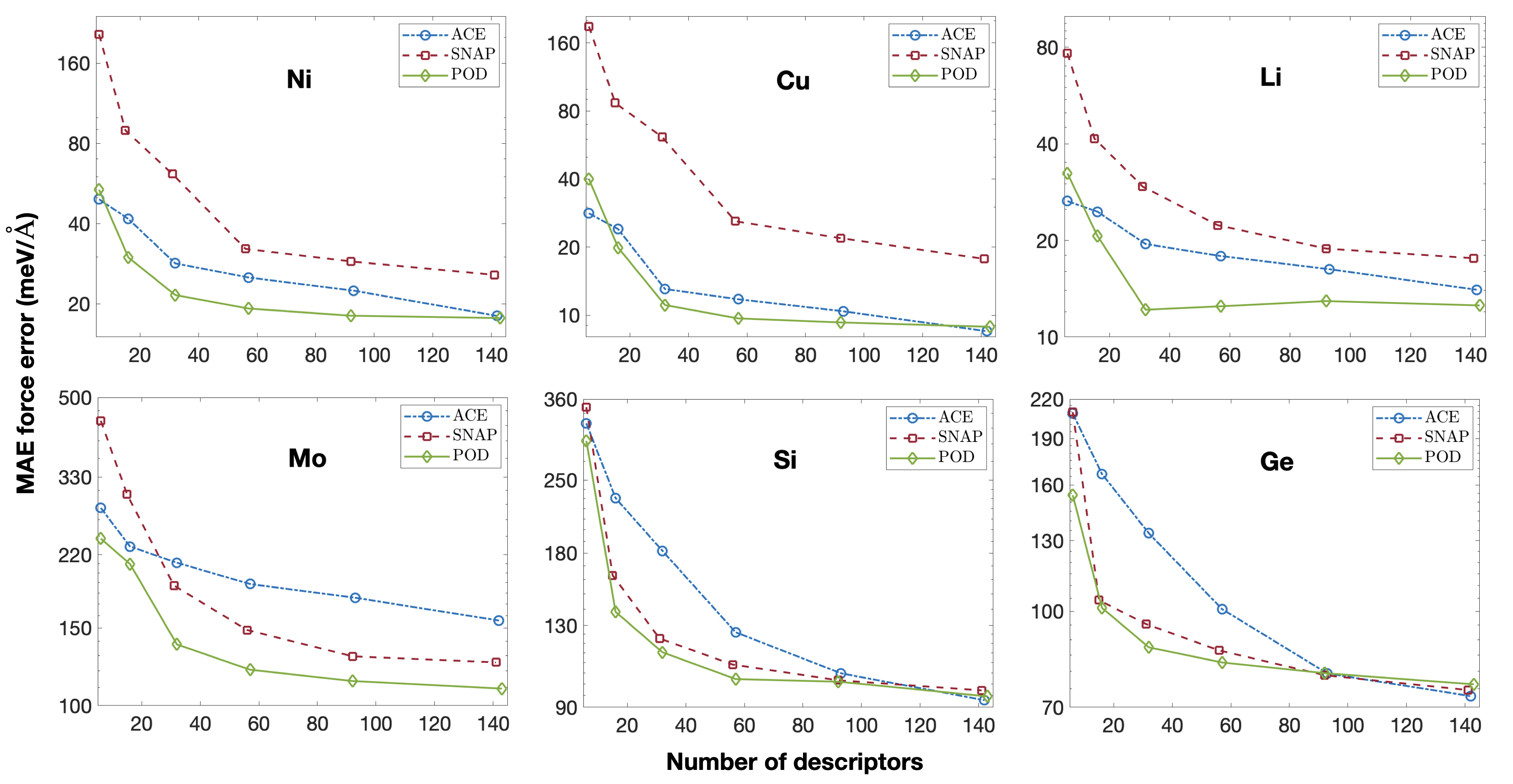}
\caption{Test errors in forces predicted using the ACE, SNAP, and POD methods as a function of the number of descriptors for Ni, Cu, Li, Mo, Si, Ge elements.}
\label{fig6}	
\end{figure}

\begin{table}[h]
\centering
\small 
	\begin{tabular}{|c|c| c c c c c c|}
		\cline{1-8}
		 Descriptors  & Data set & Ni & Cu & Li & Mo & Si & Ge \\
		\hline
		\hline 
ACE  & Training & 37.90 & 21.34 & 23.39 & 234.9 & 205.7 & 162.9 \\
$M=16$ & Test & 41.63 & 24.00 & 24.57 & 229.5 & 230.5 & 166.5 \\
 \hline
SNAP & Training & 71.98 & 71.09 & 38.96 & 306.6 & 146.2 & 102.1 \\
$M = 15$ & Test & 89.48 & 86.80 & 41.55 & 301.2 & 162.9 & 104.2 \\
\hline 
POD & Training & 26.91 & 16.46 & 19.12 & 211.1 & 127.8 & 99.90 \\
 $M=16$ & Test & 29.84 & 19.88 & 20.63 & 209.4 & 138.2 & 101.2 \\
 		\hline
        \hline 
ACE  & Training & 21.93 & 10.39 & 16.94 & 185.8 & 115.2 & 97.38 \\
$M=57$ & Test & 25.10 & 11.75 & 17.87 & 188.6 & 126.1 & 100.8 \\
 \hline
SNAP & Training & 27.23 & 22.82 & 20.65 & 143.9 & 98.77 & 85.17 \\
$M = 56$ & Test & 32.14 & 26.02 & 22.30 & 148.3 & 109.0 & 86.48 \\
\hline 
POD & Training & 16.73 & 8.355 & 11.80 & 120.3 & 91.67 & 80.06 \\
 $M=57$ & Test & 19.18 & 9.671 & 12.46 & 120.6 & 102.2 & 82.71 \\
 		\hline 	
        \hline 
ACE  & Training & 15.85 & 7.577 & 12.99 & 149.4 & 81.72 & 69.32 \\
$M=142$ & Test & 18.01 & 8.488 & 14.04 & 155.9 & 92.94 & 72.97 \\
 \hline
SNAP & Training & 21.44 & 14.69 & 16.11 & 118.8 & 85.78 & 74.69 \\
$M = 141$ & Test & 25.66 & 17.73 & 17.60 & 125.3 & 97.11 & 74.63 \\
\hline 
POD & Training & 15.01 & 7.676 & 11.61 & 107.1 & 82.94 & 72.71 \\
 $M=143$ & Test & 17.67 & 8.875 & 12.54 & 109.2 & 94.54 & 76.25 \\
 		\hline 		
	\end{tabular}
	\caption{Mean absolute errors in forces (meV/\AA) predicted using the ACE, SNAP, POD methods for the training and test sets of Ni, Cu, Li, Mo, Si, Ge elements.} 
	\label{tab5}
\end{table}

Figure \ref{fig6} depicts the force errors as a function of the number of the descriptors for the test sets, while Table \ref{tab5}  tabulates the force errors for three different values of the number of descriptors for both the training and test sets. We note that the POD potentials have smaller force errors than the ACE and SNAP potentials across all elements for $M \le 60$. In general, POD outperforms both ACE and SNAP when the number of descriptors is smaller than 60. However, ACE and SNAP achieve similar errors to POD when the number of descriptors is large enough. Again, all the potentials exhibit low test errors on forces (on the order of 10 meV/\AA) for the Ni, Cu, and Li datasets. Significantly higher force errors (on the order of 100 meV/\AA) are observed for the Mo, Si, and Ge datasets.

In terms of chemistries, we find that the lowest MAEs in energies and forces are observed for  fcc crystal structures (Ni and Cu elements), followed by bcc structures (Li and Mo elements), and that the highest MAEs are observed for  diamond structures (Si and Ge elements). These observations are consistent with the results reported in \cite{Zuo2020}.

\section{Conclusion}

We have introduced a new method called proper orthogonal descriptors (POD) to generate invariant descriptors that are appropriate for fitting potential energy surfaces. The method is based on the KL expansion of the radial and angular components of the parametrized potentials. In our work, the parametrized potentials are designed to  provide a rich and diverse representation of two-body and three-body interactions. Owing to the exponential convergence of the KL expansion, the number of descriptors  needed to reach convergence in energy and force errors is often small. 

We have evaluated  the performance of POD in light of recently developed state-of-the-art methods such as SNAP and ACE descriptors for many different elements spanning different crystal structures (fcc, bcc, and diamond), chemistries (main group
metals, transition metals and semiconductors) and bonding (metallic and covalent). This evaluation is carried out on the basis of mean absolute errors of the predicted energies and forces versus the number
of descriptors. 

For all the three methods, increasing the number of descriptors  leads to higher accuracy in predicting energies and forces. The convergence in energy and force prediction is fast as the number of descriptors increases from 10 to 60. When the number of descriptors is above 60, the convergence appears somewhat slower.  We observe that the errors for Ni, Cu, and Li elements are considerably smaller than those for Mo, Si, and Ge elements. For Ta element, the MAE in energies reaches 3.21, 1.27, and 1.07 meV/atom for ACE, SNAP, and POD, respectively. POD exhibits smaller MAEs in energies and forces than ACE and SNAP in a majority of cases.  Nonetheless, we find that all methods are able to achieve near-DFT accuracy in predicting energies and forces for all elements studied. 
 
We find that the performance of a fitted potential can be improved by optimizing the hyperparameters. To this end, we have formulated a bilevel minimization problem in which the actual loss function is replaced with a polynomial interpolation surrogate. The gradient descent algorithm is used with randomly sampled initial guesses to minimize the surrogate loss function to obtain the hyperparameters and the expansion coefficients.

The ideas presented in this paper can be extended to both empirical and machine learning potentials. The two-body orthogonal basis functions developed herein can be used as radial basis functions in power spectrum, bispectrum, SOAP and ACE descriptors. Moreover, instead of fitting the parameters of empirical potentials, one can use the proposed method to construct an orthogonal basis for the parametric manifold of empirical potentials. We believe that this approach can improve not only the accuracy but also the transferability of many existing empirical potentials.

\section*{Acknowledgements} \label{}
We would like to thank all members of the CESMIX center at MIT for fruitful and invaluable discussions leading to the ideas presented in this work. We want to thank Jaime Peraire, Youssef Marzouk, Nicolas Hadjiconstantinou, Dionysios Sema, William Moses, Jayanth  Mohan, Dallas Foster, Valentin Churavy, Mathew Swisher for the many discussions we have on a wide ranging of different topics related to this work. We would also like to thank  Christoph Ortner for making the ACE package available  and Aidan Thompson for discussions about the SNAP potential. We  gratefully acknowledge the United States  Department of Energy under contract DE-NA0003965 for supporting this work. 


\appendix

\section{Karhunen-Lo\`eve Expansion}
\label{sec::appendix}

We describe the KL expansion to generate an orthogonal basis set $\{\phi_n(\bm x)\}_{n=1}^N$ from a given family of snapshots $\{\zeta_k (\bm x)\}_{k=1}^K$ \cite{Nguyen2008a}. 
First, a two-point spatial correlation function is defined as
\begin{equation}
\mathcal{K}(\bm x, \bm x') = \frac{1}{K} \sum_{k=1}^K \zeta_k(\bm x) \zeta_k(\bm x') \ ,
\label{eq3a:1}
\end{equation}
which accepts the following spectral decomposition
\begin{equation}
\mathcal{K}(\bm x,\bm x') = \sum_{k=1}^K \lambda_k \phi_k(\bm x) \phi_k(\bm x') \ .
\label{eq3a:2}
\end{equation}
Here the set of basis functions $\phi_k(\bm x), 1 \le k \le K,$ are ordered such that the associated eigenvalues
\begin{equation}
\lambda_k =  \frac{1}{K} \sum_{l=1}^K (\phi_k(\bm x), \zeta_l (\bm x))^2
\label{eq3a:3}
\end{equation}
satisfy $\lambda_k \ge \lambda_{k+1}$, where $(\cdot, \cdot)$ denotes an appropriate inner product of two functions.

Next, for a given $N \le K$, the KL procedure consists in finding $\phi_n, 1 \le n \le N,$ so as to maximize the captured energy
\begin{equation}
\max  e_N = \sum_{n=1}^N \Big( \frac{1}{K}\sum_{k=1}^K \left(\phi_n(\bm x), \zeta_k (\bm x)\right)^2\Big) = \sum_{n=1}^N \lambda_n \ ,
\label{eq3a:4}
\end{equation}
subject to the constraints $(\phi_n (\bm x),\phi_{n'}(\bm x))=\delta_{n n'}, 1 \le n, n' \le N$. The first few basis functions thus represent the main energy-containing structures in the snapshots, with their relative importance quantified by $\lambda_n$. It can be shown that the problem~(\ref{eq3a:4}) amounts to solve the eigenfunction equation 
\begin{equation}
(\mathcal{K}(\bm x, \bm x'), \phi(\bm x')) = \lambda \phi(\bm x)
\label{eq3a:5}
\end{equation}
for the first $N$ eigenfunctions. 

The method of snapshots~\cite{sirovich87:_turbul_dynam_coher_struc_part} expresses a typical empirical eigenfunction $\phi(x)$ as a linear combination of the $\zeta_k$
\begin{equation}
\phi(\bm x) = \sum_{k=1}^K a_k \zeta_k(\bm x) \ .
\label{eq3a:6}
\end{equation}
Inserting this representation and~(\ref{eq3a:1}) into~(\ref{eq3a:5}), we immediately obtain
\begin{equation}
C a = \lambda a \ ,
\label{eq3a:7}
\end{equation}
where $C \in \mathbb{R}^{K \times K}$ is given by $C_{ij}  = \frac{1}{K} \left(\zeta_i (\bm x),\zeta_j (\bm x)\right), 1 \le i, j \le K$. The eigenproblem~(\ref{eq3a:7}) can then be solved for the first $N$ eigenvectors from which the KL basis functions $\phi_n, 1 \le n \le N,$ are constructed by~(\ref{eq3a:6}).

The optimality of the KL basis can be shown by considering an approximate representation $\hat{\zeta}_k(\bm x) = \sum_{n=1}^N \gamma_{N \, n}^k \varphi_n(\bm x)$ of $\zeta_k(x)$ for an arbitrary set of orthonormal basis functions, $\{\varphi_n\}_{n=1}^N$, and demonstrating that the KL basis is a minimizer of the error minimization problem
\begin{equation}
\min \sum_{k=1}^{K} \Big\|\zeta_k(\bm x) -  \sum_{n=1}^N \gamma_{N \, n}^k \varphi_n(\bm x)\Big\|^2 \ .
\label{eq3a:8}
\end{equation} 
Indeed, this minimization problem is equivalent to the maximization problem~(\ref{eq3a:4}), which in turn asserts the optimality of $\{\phi_n(\bm x)\}_{n=1}^N$. Furthermore, the average least-squares error can be calculated as 
\begin{equation}
\frac{1}{K}\sum_{k=1}^K \Big\|\zeta_k(\bm x) -  \sum_{n=1}^N (\phi_n(\bm x), \zeta_k(\bm x)) \phi_n \Big\|^2 = \sum_{j=N+1}^K \lambda_j \ .
\label{eq3a:9}
\end{equation} 
Expression~(\ref{eq3a:9}) gives us an idea for choosing $N$ as the smallest integer such that $\sum_{n=1}^N \lambda_n / \sum_{k=1}^{K} \lambda_k \ge 0.9999$.

The KL expansion is closely related to the proper orthogonal decomposition (POD) \cite{willcox02:_balanced_pod} and indeed equivalent to the POD when the snapshots $\xi_k$ are finite-size vectors instead of functions. In this connection, the POD can be viewed as a discrete version of the KL expansion. 

 \bibliographystyle{elsarticle-num} 
\bibliography{library}





\end{document}